\RequirePackage{fix-cm}
\documentclass[smallcondensed]{svjour3}     
\smartqed  
\usepackage{amsmath}
 \usepackage{amsfonts}
 \usepackage{mathtools}
 \usepackage{hyperref} 
 \usepackage{url}  
 \usepackage{float} 
 \usepackage{subfig}
 \usepackage{graphicx,caption} 
 \usepackage{color}
 \usepackage{dsfont}
\usepackage{algorithmicx}
\usepackage[ruled]{algorithm}
\usepackage[noend]{algpseudocode}
\usepackage{footnote}
\makesavenoteenv{table}

\begin{document}

\title{ Continuous-Time User Modeling In the Presence of Badges: A Probabilistic Approach
}


\author{Ali Khodadadi, Seyed Abbas Hosseini, Erfan Tavakoli, Hamid R. Rabiee
}


\institute{Ali Khodadadi\at
	 AICT Innovation Center, \\
              Department of Computer Engineering
              \\ Sharif University of Technology, Tehran, Iran \\     
              \email{khodadadi@ce.sharif.edu} 
           \and \\ 
           Seyed Abbas Hosseini \at
           AICT Innovation Center, \\
              Department of Computer Engineering
              \\ Sharif University of Technology, Tehran, Iran \\     
            \email{a\_hosseini@ce.sharif.edu}           
           \and \\ 
           Erfan Tavakoli \at
           AICT Innovation Center, \\
              Department of Computer Engineering
              \\ Sharif University of Technology, Tehran, Iran \\     
            \email{etavakoli@ce.sharif.edu}           
              \and \\
              Hamid R. Rabiee \at
              AICT Innovation Center, \\
              Department of Computer Engineering
              \\ Sharif University of Technology, Tehran, Iran \\     
              \email{rabiee@sharif.edu}           
}

\date{Received: \today / Accepted: \today}

\maketitle

\begin{abstract}
User modeling plays an important role in delivering customized web services to the users and improving their engagement. However, most user models in the literature do not explicitly consider the temporal behavior of users. More recently, continuous-time user modeling has gained considerable attention and many user behavior models have been proposed based on temporal point processes. However, typical point process based models often considered the impact of peer influence and content on the user participation and neglected other factors. Gamification elements, are among those factors that are neglected, while they have a strong impact on user participation in online services.  In this paper, we propose interdependent multi-dimensional temporal point processes that capture the impact of badges on user participation besides the peer influence and content factors. We extend the proposed processes to model user actions over the community based question and answering websites, and propose an inference algorithm based on Variational-EM that can efficiently learn the model parameters. Extensive experiments on both synthetic and real data gathered from Stack Overflow show that our inference algorithm learns the parameters efficiently and the proposed method can better predict the user behavior compared to the alternatives.

\end{abstract}

\section{Introduction}
\label{intro}
In recent years, online social media have become more and more popular. Social media systems such as Facebook, Twitter, Stack Overflow, and Wikipedia have millions of users engaged in various activities. In these systems, the value is created from voluntary user contributions and they need to engage users as much as possible. 
Hence, it is of key importance for these systems to provide customized services to satisfy their users and preventing user churns. Therefore, these systems need to maintain and analyze user profiles that contain a summary of important personal information. As much as the systems can create more detailed user profiles, they can provide more customized services to the users. Traditionally, the services create profiles based on the user provided information, that has two main drawbacks. First, many of the users do not create profiles and second, user activities over websites are dynamic over time. Therefore, having automated user profiling systems that consider the temporal dynamics of user activities will results in more realistic and applicable profiles. 

The unprecedented traces of user activities over social computing services provides rich information about what is done by which entities, corresponding to their location and time. These information create new opportunities for learning user interests and modeling users behavior over these systems, which can be used for creating automated user profiles. The temporal dynamics of user behavior carry a great deal of information and can help the service providers in many aspects, such as designing efficient marketing strategies, providing customized services, and preventing users' churn. For example, in an on-line shopping service it is important to predict when a user will use an item, or in a community based question and answering (CQA) site, predicting when a question will be answered is of great importance. Furthermore,  in an e-commerce service, analyzing the dynamics of service usage by the users will also help in finding potential churning users and providing them with more incentives to avoid the churn. Hence, it is important having behavior models that not only predict the type of user actions, but also model the time of user activities and their temporal dynamics. The availability of such data over time enable us to learn the key dynamic characteristics of users such as their interests, their loyalty to the provided services, and their satisfaction over time.

There exists a rich literature on modeling user activity that only considers the type of activities but does not pay attention to their timing. Indeed, these models have only concentrated on \textit{what} is done on the social media, and do not consider \textit{when} it is done \cite{abdel2013survey}. However, a large and growing interest has been emerged on modeling user behavior over time. Most of these methods do not model the time of user actions and only consider time as a covariate and hence are unable to predict the time of users future actions \cite{ahmed2013scalabledynamic,Yin2015}. Moreover, most of the existing temporal user modeling methods approach the problem in a discrete time manner. These models suffer from two main drawbacks. First, they assume the process of generating data proceeds by unit time-steps, and hence are unable to capture the heterogeneity of the time to predict the timing of the next event. Second, these models need to choose the length of time-steps to discretize the time, which is a challenging task. Actually, the precise time interval or the exact distance between two actions carries a great deal of information about the dynamics of the underlying systems and hence the emerging continuous-time models that consider the exact timing of events are more applicable to real world scenarios \cite{yang2013mixture,zhou2013learning,zarezade2015correlated,valera2015modeling,HNP3,du2015dirichlet,du2013uncover,he2015hawkestopic}.

Different factors  affect the user participation in a social computing service over time, and considering them in a continuous time user behavior model will results in more realistic models. These factors include:
\begin{itemize}
\item[$\bullet$]{\textbf{Peer Influence:}}
 The user engagement may be affected by other users activities in the social media, known as \emph{peer influence}. For example, in social networks such as Twitter and Foursquare, the user engagement in the service heavily depends on his friends activities.
\item[$\bullet$]{\textbf{Content:}}
One of the main factors that affect user participation in online services, is the content being shared over these media. As much as the content is related to the user interests, the user will be more engaged. For example, in knowledge sharing services such as Stack Overflow, the amount of user engagement depends on the content being generated over this media. 
\item[$\bullet$]{\textbf{Incentives:}}
To engage users in more contributions and steer user behaviors, many social computing services incorporate some techniques known as \emph{Gamification}. Gamification is defined as using game elements in non-game systems to increase user engagement \cite{deterding2011game}. One of the most widely used gamification elements are \emph{Badges}. Badges are some medals which are awarded to users based on some predefined levels of engagement. Recently, many social media sites are using badges to encourage users for more contributions. For example, Stack Overflow, which is a well-known questions answering service, uses badges to encourage users for more contributions. Foursquare, which is a location-based social network awards badges for user check-ins. Tripadvisor awards badges for writing reviews about different places. Recent studies have shown that these gamification elements act as \textit{incentive} mechanisms and have a significant effect on user participation behavior over social media websites \cite{anderson2013Steering}.
\end{itemize}
 
Continues-Time modeling of user activities over social media has gained considerable attention in recent years. Some preliminary works tried to model the user behavior and predict the future user actions using the time of users' actions and the impact of activity of friends on each other \cite{yang2013mixture,zhou2013learning}.
More recently, some works tried to jointly model the content and time of actions \cite{HNP3,du2015dirichlet,du2013uncover}. The main drawback of the aforementioned methods is that they only consider the impact of friends actions on user activity and ignore other factors such as gamification elements, while considering the ways in which the gamification elements impact user activities and participation in the site, is very important in user modeling.

In this paper, we propose novel multi-dimensional temporal point processes to jointly model the time and content of user actions by considering gamification elements, specially badges. We customize the method for modeling user actions over CQA sites. To this end, we propose intertwined point processes to jointly model two main type of user actions over these sites: Asking questions, Answering to the questions. Considering all the aforementioned factors in user model, the proposed method is able to infer user interests and predict the time and content of future actions. In summary, the major contributions of the proposed method are as follows:
\begin{itemize}
\item [$\bullet$]We propose a continuous-time user behavior model, that models the user activities in presence of gamification elements, especially badges. The method \textbf{Learns} the impact of badges on each user which allows us to differentiate between users.
\item[$\bullet$] We customized the method for modeling the time and content of user activities over CQA sites. We consider two main activities over these sites: asking questions, and answering questions. To do that, we define two interdependent point processes for each user. One for asking questions, and the other for answering the questions. 
\item[$\bullet$] We consider the  temporal dynamics of user actions in the proposed model. In our model, the time of current user actions depends on the content of previous actions and also the content of current user actions depends on the time of user actions.
\item[$\bullet$]  We model the dependency of the processes for asking and answering questions, using interdependent processes. Using this novel framework, beside modeling the dependency between content and time, we also model the dependency between the two different multi-dimensional processes for asking and answering questions.
\item[$\bullet$]  We propose an inference algorithm based on variational expectation maximization. The proposed inference algorithm, efficiently learns  the parameters of the model.
\end{itemize}
We evaluate the proposed method over synthetic and real datasets. The real dataset is gathered from Stack Overflow which is a popular question answering website. The results show the efficiency of proposed method in both estimating the time and content of user actions.

The remainder of this paper is organized as follows. In section \ref{sec:PriorWorks} we briefly review the relevant related works in continuous-time modeling of user activities and impact of gamification on user behavior. Details of the proposed method is discussed in section \ref{sec:proposed_model}. To demonstrate the effectiveness of the proposed model, extensive experimental results are reported and analyzed in section \ref{sec:experiments}. Finally, section \ref{sec:conclusion} concludes this paper and discusses the paths for future research.

\section{Continuous-Time User Modeling}\label{sec:PriorWorks}
The abundant available data over social media creates new opportunities for learning models of user behavior. These models can be used to predict future activity, and identify temporal information.
There exists a rich literature that use user models over social media ranging from recommender systems \cite{ahmed2011scalable,liang2016modeling}, to diffusion network analysis \cite{rodriguez2011uncovering,zhou2013learning},  and content analysis \cite{ahmed2012modeling} \cite{abdel2013survey}.
 Most of the primary works on user modeling have not paid attention to the time of activities. Some have focussed on describing aggregate behavior of many people \cite{lerman2012using,iribarren2009impact,castellano2009statistical} while others have focussed on individual behavior models \cite{yang2015modeling,hogg2013stochastic}. In this direction, the works have mainly concentrated on the impact of social network \cite{myers2012information} and  topics of interests on user activities \cite{ahmed2012modeling,yang2015modeling,chua2013generative}.
However, a large and growing literature have been emerged on modeling user behavior over time. Most of these methods do not model the time of user actions and only consider time as a covariate, and hence are unable to predict the time of users future actions \cite{ahmed2013scalabledynamic,Yin2015}. 
Moreover, most of existing temporal user modeling methods approach the problem in a discrete time manner, and the classic varying-order Markov models \cite{begleiter2004prediction,rabiner1989tutorial} have been used by most of those methods \cite{Manavoglu2003,Raghavan2014,Gao2015understanding}. These models suffer from two main drawbacks. First, they assume the process of generating data proceeds by unit time-steps, and hence are unable to capture the heterogeneity of the time to predict the timing of the next events. Second, these models need to choose the length of time-steps to discretize the time which is a challenging task. Furthermore, in the varying-order Markov models when the number of states is large, due to the exponential growth of the state-space, they can not capture the long dependency on the history of events. 

Actually, the precise time interval or the exact distance between two actions carries significant information about the dynamics of the underlying systems, and hence it is of great importance to have continuous-time models that consider the exact timing of events. Temporal point processes are a general mathematical framework for modeling continuous-time events \cite{Kingman1992,rasmussen2011temporal}. Recently, they have attracted considerable attention for modeling user activity over social media. 
 Several models have been proposed in the literature that use temporal point processes to model the information diffusion over networks \cite{yang2013mixture,zhou2013learning}. 
The first studies considered the impact of peer influence on user diffusion behavior and used a special point process, called Hawkes \cite{zhou2013learning,yang2013mixture,rodriguez2011uncovering,du2012learning}, to model the user activities. In this direction, other extensions were considered, that removed the independence assumption between different cascades \cite{zarezade2015correlated,valera2015modeling}. Some other works tried to consider the content in the diffusion behavior \cite{HNP3,du2015dirichlet,du2013uncover,he2015hawkestopic}. More recently, there have been some studies that tried to use temporal point processes in other domains. The authors in \cite{du2015time,gopalan2015scalable} incorporated temporal point processes in recommender systems. 
Linderman et al. \cite{linderman2014discovering} used point processes to infer latent network structure in financial economic interactions and reciprocity in gang violence.
Farajtabar et. al. \cite{farajtabar2015coevolve} tried to use temporal point processes to jointly model the network evolution and diffusion together, and Zarezade et. al.\cite{ZarezadeSTP16} tried to model user check-in behavior in location based social networks using temporal point processes. Our proposed model is different from the previous literature in different aspects; First, the continuous-time models in the literature mainly consider the impact of social network on the user participation and less attention is paid to the content, and more importantly, no attention is paid to the gamification elements and their impact on user participation. We extend the previous works by considering the impact of gamification elements on user participation. Second, the previous works mainly concentrate on modeling information diffusion and pay little attention to other applications. We extended the application of temporal point processes to modeling user activity over CQA sites which is substantially different from information diffusion.

Since we consider the impact of gamification elements on user actions in the proposed model, in the remainder of this section we briefly review the works on user modeling that is related to gamification.
The study of gamification and impact of gamification elements on user participation in online media has gained a considerable attention in recent years \cite{Deterding:2011Gamification}.
Most of the works done in this domain are empirical studies of user participation in social computing systems.  Antin and Churchill \cite{antin2011badges} discuss the various functions and motivations of badges in terms of their psychological incentives. Halavais et al. \cite{halavais2014badges} study the impact of social influence on individual badge earnings on Stack Overflow, and conclude that the influence of friends on badge selection is weak but has some effect.  Zhang et al. \cite{zhang2016social} study an existing badge system in Foursquare and unlike Halavais, they found that users who are friends are more likely to obtain common badges.
The authors in \cite{jin2015quick} analyze how much gamification techniques influence the member response tendencies. The authors in \cite{sinha2015modeling} study the impact of reputation on user activities and clustered users based on the reputation trends. Analyses of Stack Exchange reputation schema and its influence on user
behavior has been performed by Bosu et al. \cite{bosu2013building} and Movshovitz-Attias et al. \cite{movshovitz2013analysis}.
The authors in \cite{cavusoglu2015can}, study the impact of a hierarchical badge system on user participation and engagement at Stack Overflow. 
Their initial results present strong empirical evidence that confirms the value of the badges and their effectiveness on stimulating voluntary participation.

Besides the aforementioned works on empirical analysis of user traces of activities, some works have tried to actively study the impact of gamification on real systems. Anderson et al. \cite{anderson2014engaging} studied a large-scale deployment of badges as incentives for engagement in a massive open online course (MOOC) system. They found that making badges more salient, increases the forum engagement. Hamari \cite{hamari2015badges} actively studied the impact of gamification elements (badges) on an international peer-to-peer trading service. His results show that users in the gamified condition were significantly more likely to use the service in a more active way. While there are many studies about empirical analysis of impact of gamification on user participation, there are little works on modeling user activities in presence of badges. The main work in this domain is the seminal work of Anderson et al. \cite{anderson2013Steering}.  They analyzed the impact of badges on user behavior on Stack Overflow. They observed that as users approach the badges boundaries they steer their efforts towards achieving the badges. Using these observations they proposed a theoretical discrete-time user behavior model and evaluated it through different experiments. Marder \cite{marder2015stack} also proposed a discrete-time model of user actions and performed a regression analysis of user activities over Stack Overflow which conforms with previous empirical observations. In summary, many works have been done on empirical analysis of gamification impact on user behaviors, and little attention is paid to modeling user behavior in presence of gamification elements. Moreover, the existing user activities models, approach the problem in a discrete-time manner and do not pay any attention to the content. Our work is the first to offer a continuous-time user model which considers the impact of gamification elements on user activities, besides the content and social influence factors.
\section{Proposed Method}\label{sec:proposed_model}
In this section, we introduce the proposed method for continuous-time user modeling in presence of badges. Our model considers all the  aforementioned factors in user modeling with more emphasis on badges. The CQA websites, which have gained a considerable attention during recent years, incorporate different elements to increase user engagement. For example, Stack Overflow which is the most popular CQA website on programming questions, uses badges in an effective manner to increase user participation. Moreover, user engagement over this site depends on other users participation and the content being shared. 
These facts makes CQA websites a good candidate for continuous-time user modeling in presence of badges. Hence, we customize our model for user actions over Stack Overflow.

Asking questions and answering to others questions are the two main type of user actions over CQA sites that guarantees the survival of these sites. Therefore,  it is of great importance that the proposed method be able to model and predict users' actions in these two areas. 
In this section, we propose two interdependent multi-dimensional temporal point processes to model the asking questions and answering behaviors over a CQA site in presence of badges. In order to model the dependency of these two types of activities, we interrelate the processes to each other. In order to model the impact of badges in user participation, we develop a rich set of flexible temporal kernels that accumulate the impact of badges on the intensity function of user activity over time. Moreover, we consider the temporal dynamics in the content of user actions. Finally, we propose an efficient inference algorithm to infer the parameters of the proposed model. 

Since our model is based on stochastic temporal point processes, to make the presentation self-sufficient, some theoretical background on these processes is provided in Section \ref{subsec:Background}. The proposed generative model is described in Section \ref{subsec:proposedGenerativeModel}, followed by the details of the inference algorithm in Section \ref{subsec:inference}.
 
\subsection{Background}\label{subsec:Background}
A temporal point process is a powerful mathematical tool for modeling random events over time. More formally, a temporal point process is a stochastic process whose realizations consists of a list of time-stamped events $\{t_1, t_2, \ldots, t_n\}$ with $t_i\in \mathbb{R}^+$. Different types of activities over a CQA site, such as asking a question, and answering the questions can be considered as events generated by a point process. 

The length of the time interval between successive events is referred to as the inter-event duration. A temporal point process can be completely specified by distribution of its inter-event durations \cite{Daley2002}. Let $\mathcal{H}_t$ denote the history of events up to time $t$, then by applying the chain rule we have:	
\begin{align}\label{eqn:CompactLikelihood}
f(t_1, \ldots, t_n) = \prod_{i=1}^n f(t_i | t_1, \ldots, t_{i-1}) = \prod_{i=1}^n f(t_i|\mathcal{H}_{t_i})  
\end{align}
Therefore, to specify a point process, it suffices to define  $f^*(t)=f(t|\mathcal{H}_t )$, which is the conditional density function of an event occurring at time $t$ given the history of events.

A temporal point process can also be defined in terms of counting process $N(t)$ which denotes the number of events up to time $t$. The increment of the process, $dN (t)$, in an infinitesimal window $[t, t + dt)$, is parametrized by the conditional intensity function $\lambda^*(t)$.  The function $\lambda^*(t)$ is formally defined as the expected rate of events occurring at time $t$ given the history of events, that is:
\begin{align}
	\lambda^*(t)dt = \mathbb{E}[dN(t)|H_t]
\end{align}
There is a bijection between the conditional intensity function (intensity for short) and the conditional density function:
\begin{align} \label{eqn:lambda}
\lambda^*(t) = \frac{f^*(t)}{1-F^*(t)}
\end{align}
where $F^*(t)$ is the Cumulative Distribution Function (CDF) of $f^*(t)$. Using the definition of $\lambda^*(t)$ in eq.\ref{eqn:lambda}, the likelihood of a list of events ($t_1, \ldots, t_n$) which is observed during  a time window $[0,T)$, can be defined as: 
\begin{align} \label{eqn:PPLikelihood}
\mathcal{L} = \prod_{i=1}^{n} \lambda^*(t_i) \exp \left(-\int_0^T\lambda^*(s) ds \right)
\end{align}
where $n$ is the number of observed events and $T$ is the duration of observation. 
Intuitively, $\lambda^*(t)$ is the probability of an event occurring in time interval $[t,t+dt)$ given the history of events up to $t$, and it is a more intuitive way to characterize a temporal point process \cite{Aalen2008}. For example, a temporal Poisson process can be characterized as a special case of a temporal point process with a history-independent intensity function which is constant over time, i.e. $\lambda^*(t) = \lambda$ \cite{Kingman1992}. Users' actions usually exhibit complex longitudinal dependencies such as self-excitation, where a user tends to repeat what he has done recently. Such behavioral patterns can not  be characterized by using homogeneous a Poisson process, and hence more advanced temporal point processes are needed. Hawkes process is a temporal point process with a particular intensity function which is able to capture the self-excitation property. The intensity function of a Hawkes process is given by:
\begin{align}\label{eqn:hawkes}
\lambda^*(t) = \mu+\alpha g_{\omega}(t)\star dN(t)=\mu + \alpha \sum_{t_i<t} g_{\omega}(t-t_i)
\end{align}
where $\mu$ is a constant base intensity, $\alpha$ is a weighting parameter which controls the impact of previous events on the current intensity, and $g_{\omega}(t)$ is a kernel which defines the temporal impact of events on the future intensity. In the case that $g_\omega(t)$ is a decreasing function, Hawkes process produces clustered point patterns over time and hence is able to model the self-excitation property of users events. The right hand side of eq.\ref{eqn:hawkes} comes from the fact that the number of events occurred in a small window $[t,t+dt)$ is $dN(t) = \sum_{t_i\in \mathcal{H}_t}\delta(t-t_i)$, where $\delta(t)$ is a Dirac delta function. 

In many situations, we need to model the events generated by a set of dependent sources. Multi-dimensional point processes are a set of powerful tools for modeling such events. In a multi-dimensional point process, the intensity of a dimension depends on the event history of all dimensions. For example, as it was mentioned before, the users' actions over social media depends on each other and hence can't be modeled independently.  In order to model these action, we can use a multidimentional temporal point process in which each user corresponds to a dimension and the events of each user impacts the intensity function of other users. For example, the intensity of a multi-dimensional Hawkes process is given by:
\begin{align}\label{eqn:MDHawkes}
	\lambda_u(t)=\mu_u+ \sum_{v\in U} \alpha_{vu}g_{\omega}(t)\star dN_v(t)
\end{align}
where $\lambda_u$ shows the intensity of user $u$ to do an action, and $\mu_u$ is the base intensity of user $u$, and $\alpha_{vu}$ shows the influence of user $v$ on user $u$. As it is evident in eq.\ref{eqn:MDHawkes}, the intensity of user $u$ depends on history of all users through the second term. Hawkes process models the dependency among different dimensions through the convolution with a temporal kernel which is a linear operator. However, the event in different dimensions may exhibit more complicated dependencies and hence we need more complex methods for modeling such phenomena. For example, in order to model the effect of badges on user activities, we propose a nonlinear multi-dimensional temporal point process in section \ref{subsec:proposedGenerativeModel}.

Each event can also be associated with some auxiliary information known as the mark of an event. For example, the tags of the questions in a CQA website can be considered as the marks of events. A marked temporal point process is a point process for modeling such events. If $k$ denotes the mark of the events, then the intensity of the marked temporal point process is given by:
\begin{align}\label{eqn:markedLambda}
\lambda^*(t,k) = \lambda^*(t) f^*(k|t) 
\end{align}
where $\lambda^*(t)$ denotes the temporal intensity function, and $f^*(k|t)$ is the conditional probability density function of observing an event with mark $k$ at time $t$. Therefore, in order to determine a temporal point process, we need a temporal intensity which shows the rate of occurring event given the history and a conditional probability density function over marks. We propose a nonlinear multi-dimensional marked point process to model  user activities in presence of badges, in the next section.

\subsection{Proposed Generative Model}\label{subsec:proposedGenerativeModel}
In this section, we propose multi-dimensional marked point processes to model the user activities over a CQA website in the presence of badges. We consider two main type of activities over such websites, i.e. asking questions and answering them. In order to model such actions using temporal point processes, we consider each of these actions as marked events and propose two intertwine point processes to model these two set of dependent events. In the following, we first detail our notations and assumptions and then introduce the proposed generative model for user actions over CQA sites.

Let $\mathbf{D}^q(t) = \{e^q_i\}_{i=1}^{N^q(t)}$ and $\mathbf{D}^a(t) = \{e^a_i\}_{i=1}^{N^a(t)}$ denote the set of asking questions and answering events observed until time $t$, respectively. We denote the number of asking questions and answering events up to time $t$ by $N^q(t)$ and $N^a(t)$, respectively. The event $e^q_i$ is a triple $(t_i,u_i,z_i)$ which indicates that at time $t_i$, user $u_i$ asks a question with tag $z_i$. The event $e^a_i$ is also a triple $(t_i,u_i,p_i)$ which indicates that at time $t_i$, user $u_i$ answers a question with id $p_i$. In the following, we propose two dependent multi-dimensional marked temporal point process to model these two dependent set of marked events.

We can consider different intents behind user actions over a CQA site. User activities may be due to drivers external to the website which we call \emph{exogenous} activities or the influences he receives from the media, such as others actions or website incentives like badges which we call \emph{endogenous} activities. Paying attention to these drivers will result in more realistic user models. Hence, we assume that the user intensities for asking questions and answering questions are as follows:
\begin{align}\label{rel:q_intensity}
\lambda_u^q(t) = \underbrace{\vphantom{\sum_{b \in \mathcal{B}^q}}\mu_{u}^q}_{\text{Exogenous Intensity}}  +  \underbrace{\rho_u^q \sum_{b \in \mathcal{B}^q} g_w^q(h_b(D_u^q(t)), \tau_b) }_{\text{Endogenous Intensity}}
\end{align} 
\begin{align}\label{rel:a_intensity}
\lambda_u^a(t) &= \underbrace{\vphantom{\sum_{b \in \mathcal{B}^q}}\mu_{u}^a}_{\text{Exogenous Intensity}} +
\underbrace{\rho_u^a \sum_{b \in \mathcal{B}^a} g_w^a(h_b(D_u^a(t)), \tau_b) + \sum_{e_i \in D_.^q(t)} \eta_{uz_i} f_w(t,t_i)}_{\text{Endogenous Intensity}}
\end{align} 
We consider the gamification elements (more specifically; badges) as the main drivers of endogenous activities. Badges are awarded to the users based on predefined levels of engagement. Recent analysis has shown that the badges increase user participation over CQA sites. It have also been noticed that the amount of increase depends on how much the user is close to the badge; i.e. the impact of badge on users participation increases as they are close to achieving it \cite{Gao2015understanding}.
 
The selected kernels in user intensity should be able to reflect the two aforementioned facts. They should be also able to handle the heterogeneity in the criteria for wining badges. For example, some badges are awarded based on the mount of activities of a given type (we call it a badge $b_1$), while others are awarded based on number of days the user performed a given type of activity (we call it $b_2$). To do that, we define the parameter $\tau_b$ that captures the criteria for winning the badge $b$. For our examples, $\tau_{b_1}$ is the total amount of actions which is required to win a badge, while $\tau_{b_2}$ is the amount of active days required to win a badge. We also define per badge functions $h_b$ that extracts the required information from the history of user actions related to a specified badge. Again for our example badges, $h_{b_1}(D_u(t)) = N_u(t)$ , which is the count of total actions of user $u$ until time $t$, while $h_{b_2}(t)$ is the total number of active days of user $u$, until time $t$. Finally, the temporal kernels $g_w^q(x,y)$ and $g_w^a(x,y)$, represent the impact of badges on user intensity. To capture the two aforementioned facts, they should have the following features:
\begin{itemize}
\item \textbf{Non-negativity}: To capture the positive impact of badges on user participation, the kernels should never become negative. Here, we consider only a positive impact for badges on user participation. The negative impact of miss-designed badges on user participation, can be a good direction for the future research. 
\item \textbf{Exponential Increase:} To capture the observed feature of badges for which the impact of badge on users participation increases as they are close to achieving it, the kernels should be able to capture these phenomenon. 
\end{itemize}
Different kernels can be utilized that have the required features.  Gaussian RBF and Exponential kernels are among the most popular choices. We used the Gaussian RBF kernel $g_{\text{gauss}}$ defined as:
\begin{align}
g^{\text{gauss}}_\omega (x,y) = e^{-(\frac{y-x}{2\omega})^2}
\end{align}
And, exponential kernel $g_{\text{exp}}$ as:
\begin{align}
g^{\text{exp}}_\omega (x,y) = e^{-\omega(y-x)}\mathds{I}[y \geq x] 
\end{align}

\begin{figure}[!t] 
\centering
\includegraphics[width=2in]{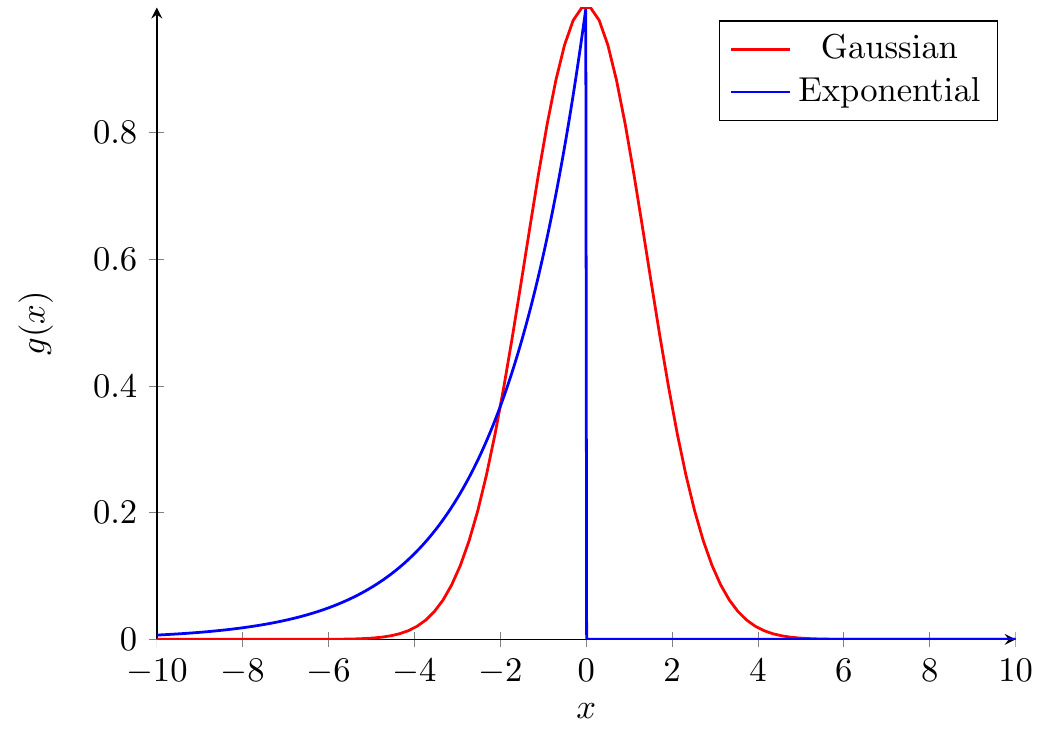}%
\caption{Different Kernels and their impact}
\label{fig:kernels}
\end{figure}
Fig. \ref{fig:kernels} shows the two selected kernels and their features. The second term in user intensities (eqs. \ref{rel:a_intensity}, \ref{rel:q_intensity}), reflects the cumulative impact of badges on users activities. To capture the heterogeneity of users, motivated by badges, we have also incorporated personal parameters $\rho_u^{a}$ and $\rho_u^{q}$. 
As mentioned before, asking questions and answering questions activities, are interwoven. Since, in a CQA site the asking question activity is mainly derived by exogenous factors, we did not consider any impact from the previous answering activities. However, we considered a positive impact of questioning activities on answering behaviors. This impact is captured through the third part of the user intensity for answering questions in eq. \ref{rel:a_intensity}. It captures the impact of other users previous questions on the user answering intensity. 

Another important factor which impacts the user activities is the content. The users usually have interest in some fields and also have some expertise in other fields. Therefore, they will ask questions and answer to questions in some limited number of fields. Hence, we considered the temporal effect of content on the intensity of users for answering questions. We modeled the content of user activities as the mark of events in the proposed temporal point processes. The mark for a question is the tag associated to it, and for an answer is the question it belongs to. The proposed mark probabilities for questions and answers are as follows:
\begin{align}\label{rel:q_mark_prob}
\mathbf{P}(z_i=k| t_i,u_i) = \frac{\alpha_{u_ik}}{\sum_{s=1}^K \alpha_{u_is}}
\end{align}
\begin{align} \label{rel:a_mark_prob}
\mathbf{P}(p_j=i| t_j,u_j) =\frac{\eta_{u_jz_i}f^a_\omega(t_i, t_j) }{\sum_{e_r \in \mathbf{D}^q_.(t_j)}\eta_{u_jz_r}f^a_\omega(t_r, t_j) } 
\end{align}
Let $\alpha_u$ be a vector of user interest over tags, and let $K$ be the total number of tags, then $\alpha_u$ is a $K$ dimensional vector, where $\alpha_{uk}$ represents the interest of user $u$ in asking questions of domain $k$. Also, the parameter $\eta_u$ is a $K$ dimensional vector showing the expertise of $u$
in different domains. Another important factor which has a great impact on the content of user actions is time. We also consider the negative impact of time on user answering, to capture the fact that users have less interest to answer older questions. The exponentially decaying function $f_w^a(t_i, t_j)$ captures the negative effect of time, and for simplicity we consider the widely usaed exponential function $f^a_w(t_i, t_j) = e^{-w(t_j - t_i)}$.

In summary, the proposed process can capture the following desirable properties:
\begin{itemize}
\item \textit{Capturing heterogeneous impact of badges:}
The proposed intensity functions for both asking questions and answering questions, captures heterogeneous impact of badges on user participation through the functions $h_b(.)$, and the decaying kernels $g_w^q(.)$ and $g_w^a(.)$.
\item \textit{Interdependency and Mutual Excitation:} 

The events of users have an impact on the others events. For example, The existence of more questions will results in more answers (the third part of user answering intensity in eq.\ref{rel:a_intensity}).

\item \textit{Impact of time on content and impact of content on time:}
The proposed model captures the impact of time on content through the proposed mark probability function in eq. \ref{rel:a_mark_prob}.
We also consider the impact of content on timing of user answers, utilizing the impact of user interests $\eta_{uk}$ on intensity function in third part of eq.\ref{rel:a_intensity}.
\item \textit{Temporal decays:}
Both, the impact of previous content on the time of events and the processes on each other exponentially decays as a function of time differefnce, through the decaying functions $f_w^a(.)$ in eqs.\ref{rel:a_mark_prob}, and \ref{rel:a_intensity}. This means the model pays more attention to recent actions.
\end{itemize}

\alglanguage{pseudocode}
\begin{algorithm}[t]
\small
\caption{Variational Expectation Maximization UMUB}
\label{alg:UMUB_inference}
\begin{algorithmic}[1]
	\For {each user $u\in U$}
		\State initialize $\rho_u^a,\rho_u^q,\mu_u^a,\mu_u^q,\eta_u$ randomly
	\EndFor
\While {$\Delta\log\mathcal{L} > \delta$}
 \Comment{\mbox{check for model convergence}}
 \State{\textbf{E Step:}}
 \For {each user $u\in U$}
	\For {each event $e^q_i\in D^q_u$}
		\State update $\phi^q_i$ using Eq.~\ref{eqn:question_s_update}
	\EndFor
	\For {each event $e^a_i\in D^a_u$}
		\State update $\phi^a_i$ using Eq.~\ref{eqn:answer_s_update}
		\State update $\zeta^a_i$ using Eq.~\ref{eq:update_zeta}
	\EndFor
\EndFor
\State{\textbf{M Step:}}
\For {each user $u\in U$}
	\State update $\mu^q_u$ using Eq.~\ref{eq:update_mu_q}
	\State update $\mu^a_u$ using Eq.~\ref{eq:update_mu_a}
	\State update $\rho^q_u$ using Eq.~\ref{eq:update_rho_q}
	\State update $\rho^a_u$ using Eq.~\ref{eq:update_rho_a}
	\State update $\alpha_u$ using Eq.~\ref{eq:update_alpha}
	\State update $\eta_u$ using Eq.~\ref{eq:update_eta}
\EndFor
\EndWhile
\end{algorithmic}
\end{algorithm}
\subsection{Inference}\label{subsec:inference}
In this section, we discuss the details and practical challenges of estimating the model parameters by using the users activity data, and present a variational expectation maximization algorithm as a scalable solution.

Given the users' activities in a time window $[0,T]$, and the data $\mathbf{D} = \mathbf{D} ^q \cup \mathbf{D} ^a $, we would like to infer the model parameters $\Theta = \{ \mu_u^q,\mu_u^a,\rho_u^q,\rho_u^a, \eta_u, \alpha_u \}_{u\in U}$. Based on the proposed generative model, the log-likelihood of observed data $\log p( \mathbf{D}|\Theta)$ is given by:
 \begin{align}
\log p(\mathbf{D}| \Theta) =\sum_{u\in U} \log p(\mathbf{D} _u | \theta_u) 
\end{align}
where $\theta_u$ is the parameter set for user $u$, i.e. $\{ \mu_u^q,\mu_u^a,\rho_u^q,\rho_u^a, \eta_u, \alpha_u \}$. Using the chain rule, we can further split the likelihood for each user as:
\begin{align}
\log p(\mathbf{D}_u| \theta_u) = \log p(\mathbf{D} ^q_u | \theta_u) + \log p(\mathbf{D} _u^a | \mathbf{D} _u^q,\theta_u)   
\end{align}
where, the first term is the log-likelihood of user $u$'s asking question activities, and the second term is the log-likelihood of answering activities of user $u$. Using the theory of point processes, i.e. eq. \ref{eqn:PPLikelihood} and eq. \ref{eqn:markedLambda}, we have:
\begin{align}\label{eq:TotalLogLikelihood}
\log p(\mathbf{D} | \Theta) &= \sum_{e_i \in \mathbf{D}^q} \log(\lambda_{u_i}^q(t_i)) +  \sum_{e_i \in \mathbf{D}^q} \log(f_{u_i}^q(z_i|t_i))- \sum_{u} \int_0^T \lambda_u^q(\tau) d\tau \\\nonumber
&+\sum_{e_j \in \mathbf{D}^a} \log(\lambda_{u_j}^a(t_j))+\sum_{e_j \in \mathbf{D}^a} \log(f_{u_j}^a(p_j|t_j)) - \sum_{u} \int_0^T \lambda_u^a(\tau) d\tau  
\end{align}

We would like to find the parameter set $\Theta$ that maximizes this log likelihood function. However, maximizing this log likelihood function turns out to be a complex problem. The difficulty arises from the summation over different components of the intensity functions that appears inside the logarithm in eq. \ref{eq:TotalLogLikelihood}:
\begin{align}
	\log(\lambda_{u_i}^q(t_i)) &= \log \left(\mu_{u_i}^q + \rho_{u_i}^q \sum_{b \in \mathcal{B}^q} g_w^q(h_b(\mathbf{D}_{u_i}^q(t_i)), \tau_b) \right) \\\nonumber
	\log(\lambda_{u_j}^a(t_j)) &= \log\left(\mu_{u_j}^a + \rho_u^a \sum_{b \in \mathcal{B}^a} g_w^a(h_b(\mathbf{D}_{u_j}^a(t_j)), \tau_b) + \sum_{e_i \in \mathbf{D}_.^q(t)} \eta_{uz_i} f^a_w(t_j,t_i) \right)
\end{align}
Therefore, if we set the derivatives of the log likelihood to zero, we would not obtain a closed form solution.
In order to resolve this issue, we first give an alternative formulation of the model in which we add an additional layer of latent variables. These auxiliary variables facilitate the inference algorithm without changing the model \cite{gopalan2015scalable}. As we mentioned in section \ref{subsec:proposedGenerativeModel}, there are different factors that trigger the users' actions. Hence, for each activity $e_{i}$ of user $u$, we introduce a latent variable $s_i$ which denotes the latent factor that trigger the action $e_{i}$. For questions in $\mathbf{D}^q$, $s_i$ is a binary random variable which denotes if the badges are the driving factor of action $e_{i}$ or the driving factor is exogenous:
\begin{align}\label{eqn:completeQuestionIntensity}
	\lambda^q_{u}(t_i,s_i)=\begin{cases}
			\mu^q_u 				&\text{$s_i=1$}\\
	\rho_{u}^q \sum_{b \in \mathcal{B}^q} g_w^q(h_b(\mathbf{D}_{u}^q(t_i)), \tau_b)	&\text{$s_i=0$}
 \end{cases}
\end{align}
For each answering activity $e_{j}$ in $\mathbf{D}^a$,  let $s_j \in \{-1,0\}\cup \{1,2,\ldots,|D^q(t_j)|\}$ be the latent factor, where $s_j=0$ indicates that the badge incentives drove the user to answer the questions, $s_j=r>0$ indicates that the correspondence between the topic of question $r$ with user $u$'s expertise triggered him to answer the question, and $s_j=-1$ shows that an external factor triggered the answer event. By using the latent variable $s_j$, the conditional intensity of user $u$ for answering questions can be written as:
\begin{align}\label{eqn:completeAnswerIntensity}
	\lambda^a_{u}(t_j,s_j)=\begin{cases}
			\mu^a_u 				&\text{$s_j=-1$}\\
	 \rho_{u}^a \sum_{b \in \mathcal{B}^a} g_w^a(h_a(\mathbf{D}_{u}^a(t_j)), \tau_b)			&\text{$s_j=0$}\\
	 \eta_{uz_{s_j}} f_w^a(t_j-t_{s_j}) &\text{$s_j\in \{1,2,\ldots,|\mathbf{D}^q(t_j)|\}$}
 \end{cases}
\end{align}
It is known that the sum of Poisson processes is itself a Poisson process with rate equal to the sum of all individual rates. Thus, these new latent variables preserve the marginal distribution of the observation. Combining eq. \ref{eqn:completeQuestionIntensity} and \ref{eqn:completeAnswerIntensity} with eq. \ref{eq:TotalLogLikelihood}, the log likelihood of observations $\mathbf{D}$ and auxiliary latent variables $\mathbf{S}$ is given by:
\begin{align}\label{eqn:completLogLikelihood}
\log p(\mathbf{D}, \mathbf{S}| \Theta) &= \sum_{e_i \in \mathbf{D}^q} \log(\lambda_{u_i}^q(t_i, s_i)) +  \sum_{e_i \in \mathbf{D}^q} \log(f_{u_i}^q(z_i|t_i))- \sum_{u} \int_0^T \lambda_u^q(\tau) d\tau \\\nonumber
&+\sum_{e_j \in \mathbf{D}^a} \log(\lambda_{u_j}^a(t_j, s_j))+\sum_{e_j \in \mathbf{D}^a} \log(f_{u_j}^a(p_j|t_j)) - \sum_{u} \int_0^T \lambda_u^a(\tau) d\tau  
\end{align}
which can be written as:
\begin{align}\label{eqn:sumLogCompleteIntensity}
	\sum_{e_i \in \mathbf{D}^q} \log(\lambda_{u_i}^q(t_i, s_i))  &= \sum_{u \in U}C^q_{u1}\log \mu_u^q+C^q_{u0}\log \rho_u^q\\\nonumber
	\sum_{e_j \in \mathbf{D}^a} \log(\lambda_{u_j}^a(t_j, s_j)) &= \sum_{u \in U}C^a_{u,-1}\log \mu_u^a+C^a_{u0}\log \rho_u^a+\sum_{u\in U}\sum_{z\in Z}{C^a_{uz}\log{\eta_{uz}}}
\end{align}
where $C^q_{u1}$ and $C^a_{u,-1}$ denote the number of times that user $u$ ask question or answer a question based on an external factor, respectively. $C^q_{u0}$ and $C^a_{u0}$  also denote the weighted number of times that the user actions are triggered by the badge incentives. $C^a_{uz}$ also denotes the weighted number of times that user $u$'s answers triggered by questions in topic $z$:
\begin{align}\label{eqn:intensityCounts}
	C^q_{u1} &= \sum_{e_i \in \mathbf{D}_u^q} \mathds{I}[s_i = 1] \\\nonumber
	C^q_{u0} &= \sum_{e_i \in \mathbf{D}_u^q} \mathds{I}[s_i = 0] \times \rho_{u}^q \sum_{b \in \mathcal{B}^q} g_w^q(h_b(\mathbf{D}_{u}^q(t_i)),\tau_b)\\\nonumber
	C^a_{u,-1} &= \sum_{e_j \in \mathbf{D}_u^a} \mathds{I}[s_j = -1]  \\\nonumber
	C^a_{u0} &= \sum_{e_j \in \mathbf{D}_u^a} \mathds{I}[s_j = 0] \times \rho_{u}^a \sum_{b \in \mathcal{B}^a} g_w^a(h_b(\mathbf{D}_{u}^a(t_j)),\tau_b)\\\nonumber
	C^a_{uz} &= \sum_{e_j \in \mathbf{D}_u^a}\sum_{e_i \in \mathbf{D}_u^q(t_j)} \mathds{I}[z_i = z]\mathds{I}[s_j = i]\times \eta_{uz}f_w^a(t_j - t_i)\\\nonumber
\end{align}  
As it is mentioned before, the direct optimization of $\log p(\mathbf{D}|\Theta)$ is difficult, but the optimization of complete-data log-likelihood $\log p(\mathbf{D},\mathbf{S}|\Theta)$ is significantly easier. For any distribution $q(\mathbf{S})$ over the auxiliary latent variables, the following decomposition holds \cite{Bishop2006PRM}:
\begin{align}
\log p(\mathbf{D}|\Theta) = \mathcal{L}(q(\mathbf{S}),\Theta)+\mathbf{KL}\left(q(\mathbf{S})||p(\mathbf{S}|\mathbf{D},\Theta)\right)
\end{align}
where the functions $\mathcal{L}(q(\mathbf{S}),\Theta)$ and $\mathbf{KL}\left(q(\mathbf{S})||p(\mathbf{S}|\mathbf{D})\right)$ are defined as follows:
\begin{align}
	  \mathcal{L}(q(\mathbf{S}),\theta)&= \mathbb{E}_q\left[ \frac{p(\mathbf{D},\mathbf{S}|\Theta)}{q(\mathbf{S})} \right]\\
	\mathbf{KL}\left(q(\mathbf{S})||p(\mathbf{S}|\mathbf{D},\Theta)\right)&= -\mathbb{E}_q\left[ \frac{p(\mathbf{S}|\mathbf{D},\Theta)}{q(\mathbf{S})} \right]
\end{align}
$\mathbf{KL}\left(q(\mathbf{S})||p(\mathbf{S}|\mathbf{D}, \Theta)\right)$ is the Kullback-Leibler divergence between $q(\mathbf{S})$ and the posterior distribution $p(\mathbf{S}|\mathbf{D}, \Theta)$, and satisfies $\mathbf{KL} \geq0$ with equality, if and only if, $q(\mathbf{S})=p(\mathbf{S}|\mathbf{D}, \Theta)$. Therefore, $\mathcal{L}(q,\theta)$ is a lower bound on the log-likelihood function. In order to find the value of $\Theta$ that maximizes the log-likelihood function, we use the Variational Expectation Maximization (Variational-EM) algorithm. Variational-EM is an iterative optimization algorithm which has two main steps in each iteration.  In the E-step, the lower bound $\mathcal{L}(q(\mathbf{S}),\Theta)$ is maximized with respect to $q(\mathbf{S})$ while holding $\Theta$ fixed. In the subsequent M-step, the distribution $q(\mathbf{S})$ is held fixed, and the lower bound $\mathcal{L}(q(\mathbf{S}),\Theta)$ is maximized with respect to $\Theta$. In the E-step, we should find the posterior of the latent variables $\mathbf{S}$ in order to maximize the $\mathcal{L}(q(\mathbf{S}),\Theta)$. However, since finding the joint posterior of all latent variables is not computationally tractable, we use the mean-filed approximation assumption over $q$:
\begin{align}
q(\mathbf{S}) = \prod_{e_i \in \mathbf{D}^q} q(s_i|\phi^q_i)\prod_{e_j \in \mathbf{D}^a} q(s_j|\phi^a_j)
\end{align}
In other words, we find the best factorized $q(S)$ distribution that is most similar to the  posterior $p(\mathbf{S}|\mathbf{D}, \Theta)$ in the KL-divergence sense. In the M-Step, we optimize the $\mathcal{L}(q(\mathbf{S}),\Theta)$ with respect to $\theta$.

Although introducing the auxiliary latent variables simplify the inference algorithm, we have another challenge in inferring the answer parameters. 
While computing the $\log f_{u_i}^a(p_i|t_i)$,
the $\log \left\lbrace \sum_{e_r \in \mathbf{D} ^q_.(t_j)} \eta_{uz_r}f_w^a( t_j -t_r) \right\rbrace$ that exists in the denominator of $f_{u_i}^a(p_i|t_i)$ makes the inference of user expertise challenging.  Using the approach used in \cite{blei2006dynamic,blei2007correlated}, we define a new variational parameters $\zeta_j$  for each answer event:
\begin{align} \label{eqn:ans_lg_sum_UP}
\log \left\lbrace \sum_{e_r \in \mathbf{D} ^q_.(t_j)} \eta_{uz_r}f_w^a(t_j - t_r) \right\rbrace 
\leq
\zeta_j \left( \sum_{e_r \in \mathbf{D} ^q_.(t_j)} \eta_{uz_r}f_w^a(t_j - t_r) \right) -1 - \log(\zeta_j)
\end{align} 

In summary, we should tighten the lower bound $\mathcal{L}(q(\mathbf{S}),\Theta)$ with respect to $q(S)$ and $\zeta$ in the E-step, and maximize it with respect to $\Theta$ in the M-Step.
\subsubsection{E-Step:}
Substituting eqs. \ref{eqn:ans_lg_sum_UP}, \ref{eqn:intensityCounts} and \ref{eqn:sumLogCompleteIntensity} in eq. \ref{eqn:intensityCounts}, and considering a multinomial distribution for each $q(s_i)$ with parameter $\phi_i$, given the current state of all parameters, 
we find $q(s_i)$ that maximizes the lower bound in the E-step, which is equivalent to updating the $\phi_i$ parameters. Using some straightforward calculations, the update equations for the question latent variables, is given by:    
\begin{align}\label{eqn:question_s_update}
	\phi^q_{ik}=q(s_i=k)=\begin{cases}
		\frac{\rho_{u}^q \sum_{b \in \mathcal{B}^q} g_w^q(h_b(\mathbf{D}_{u}^q(t_i)), \tau_b)}{\mu_u^q + \rho_{u}^q \sum_{b \in \mathcal{B}^q} g_w^q(h_b(\mathbf{D}_{u}^q(t_i)), \tau_b)}&\text{$k=0$}\\
	 	\frac{\mu_u^q}{\mu_u^q + \rho_{u}^q \sum_{b \in \mathcal{B}^q} g_w^q(h_b(\mathbf{D}_{u}^q(t_i)), \tau_b)}&\text{$k=1$}
 \end{cases}
\end{align}
and, the update equations for the answer latent variables is given by:
\begin{align}\label{eqn:answer_s_update}
	\phi^a_{jk} &= q(s_j=k) \\\nonumber 
	&=\begin{cases}
		 \frac{\mu_u^a}{\mu_u^a + \rho_{u}^a \sum_{b \in \mathcal{B}^q} g_w^a(h_b(\mathbf{D}_{u}^a(t_i)), \tau_b) +\sum_{e_i \in \mathbf{D}_.^q(t_j)} \eta_{uz_i} f_w^a(t_j-t_i)}				&\text{$k=-1$}\\
\frac{\rho_{u}^a \sum_{b \in \mathcal{B}^q} g_w^a(h_b(\mathbf{D}_{u}^a(t_i)), \tau_b)}{\mu_u^a + \rho_{u}^a \sum_{b \in \mathcal{B}^q} g_w^a(h_b(\mathbf{D}_{u}^a(t_i)), \tau_b) +\sum_{e_i \in \mathbf{D}_.^q(t_j)} \eta_{uz_i} f_w^a(t_j-t_i)}	 &\text{$k=0$}\\
	 \frac{\eta_{uz_k} f_w^a(t_j-t_k) }{\mu_u^a + \rho_{u}^a \sum_{b \in \mathcal{B}^q} g_w^a(h_b(\mathbf{D}_{u}^a(t_i)), \tau_b) +\sum_{e_i \in \mathbf{D}_.^q(t_j)} \eta_{uz_i} f_w^a(t_j-t_i)}	
	 &\text{$k\in \{1,2,\ldots,|\mathbf{D}^q(t_j)|\}$}
	 \end{cases}
\end{align}
finally, the update equations for variational parameters $\zeta_j$ is given by:
\begin{align}\label{eq:update_zeta}
\zeta_j = \frac{1}{\sum_{e_r \in \mathbf{D} ^q_.(t_j)} \eta_{uz_r}f_w^a(t_j - t_r)}
\end{align}

\subsubsection{M-Step:}
In the subsequent M-step, the distribution $q(\mathbf{S})$ is held fixed and the lower bound $\mathcal{L}(q(\mathbf{S}),\Theta)$ is maximized with respect to $\Theta$, where
$\Theta = \{ \mu_u^q,\mu_u^a,\rho_u^q,\rho_u^a, \eta_u, \alpha_u \}_{u\in U}$. 
To find the point estimations for different parameters, we should maximize the lower bound with respect to different parameters.
Maximizing the lower bound with respect to $\mu_u^q$ and $\rho_u^q$ will result in the following closed form solutions:
\begin{align}\label{eq:update_mu_q}
\hat{\mu}^q_u =& \frac{\sum_{e_i \in \mathbf{D}_u^q} \phi^q_{i0}}{T} 
\end{align}
\begin{align}\label{eq:update_rho_q}
	\hat{\rho}^q_u =& \frac{\sum_{e_i \in \mathbf{D}_u^q} \phi^q_{i1}}{\sum_{b \in \mathcal{B}^q} \mathbf{G}^q(u,b,T)}
\end{align}
where $\mathbf{G}^q(u,b,T) = \int_0^T  g_w^q(h_b(\mathbf{D}_{u}^q(s)), \tau_b)ds$,
and maximizing the lower bound with respect to $\mu_u^a$ and $\rho_u^a$ will result in:
 \begin{align}\label{eq:update_mu_a}
\hat{\mu}^a_u = \frac{\sum_{e_j \in \mathbf{D}_u^a} \phi^a_{j,-1}}{T} 
\end{align}
\begin{align}\label{eq:update_rho_a}
	\hat{\rho}^a_u =& \frac{\sum_{e_j \in \mathbf{D}_u^a} \phi^a_{j,0}}{\sum_{b \in \mathcal{B}^a} \mathbf{G}^a(u,b,T)}
\end{align}
Maximizing the lower bound with respect to $\alpha_u$ will result in the following simple closed form solution:
\begin{align}\label{eq:update_alpha}
\hat{\alpha}_{uk} = \frac{\sum_{e_i \in \mathbf{D}_u^q}\mathds{I}[z_i = k]}{|\mathbf{D}_u^q|}
\end{align}
Finally, to find $\eta_u$ we should maximize the lower bound with respect to $\eta_u$. Unfortunately, there is no closed form solution for $\eta_u$, and we should solve the following optimization to estimate the vector $\eta_u$:
\begin{align} \label{eq:update_eta}
\hat{\eta}_u = \arg \max _{\eta_{u}} & \quad \log \eta_{u}^TF_u - \eta_{u}^TH_u \nonumber \\
s.t. & \sum_{k=1}^K \eta_{uk}=1
\end{align}
where, $F_u$ and $H_u$ are $K$ dimensional vectors defined over different tags, and for each tag $k$ we have:
\begin{align}
F_u(k) =& \sum_{e_j \in \mathbf{D}_u^a}  \left\lbrace \mathbb{I}(z_{p_j}=k) +\sum_{e_i \in \mathbf{D}_.^q(t_j), z_i = k} \phi^a_{ji} \right\rbrace  \\
H_u(k) =& \sum_{e_i \in \mathbf{D}_.^q, z_i = k} \frac{1}{\beta} \left[ 1-e^{-\beta (T-t_i)} \right] \\
+& \sum_{e_j \in \mathbf{D}_u^a} \zeta_j \left( \sum_{e_r \in \mathbf{D} ^q_.(t_j), z_r = k} f^a_w( t_j-t_r) \right)
\end{align}
Since, eq. \ref{eq:update_eta} is convex in $\eta_u$, we can find the optimal solution by using different convex optimization tools such as CVX. The overall steps of inference algorithm is depicted in algorithm \ref{alg:UMUB_inference}.

\section{Experiments}\label{sec:experiments}

\subsection{Dataset Description}
In this section, we evaluate the proposed inference algorithm, and the effectiveness of the proposed method by performing several experiments on synthetic and real datasets. To validate the proposed inference algorithm, we generate a set of events by using the proposed method, and validate the estimated parameters by using different criteria. Moreover, to evaluate the performance of our model, we use a real dataset which has been collected by crawling the Stack Overflow. The datasets are explained in more details in the following paragraphs.
\begin{figure*}[!t]
\centering
\subfloat[]{\includegraphics[width=2.2in]{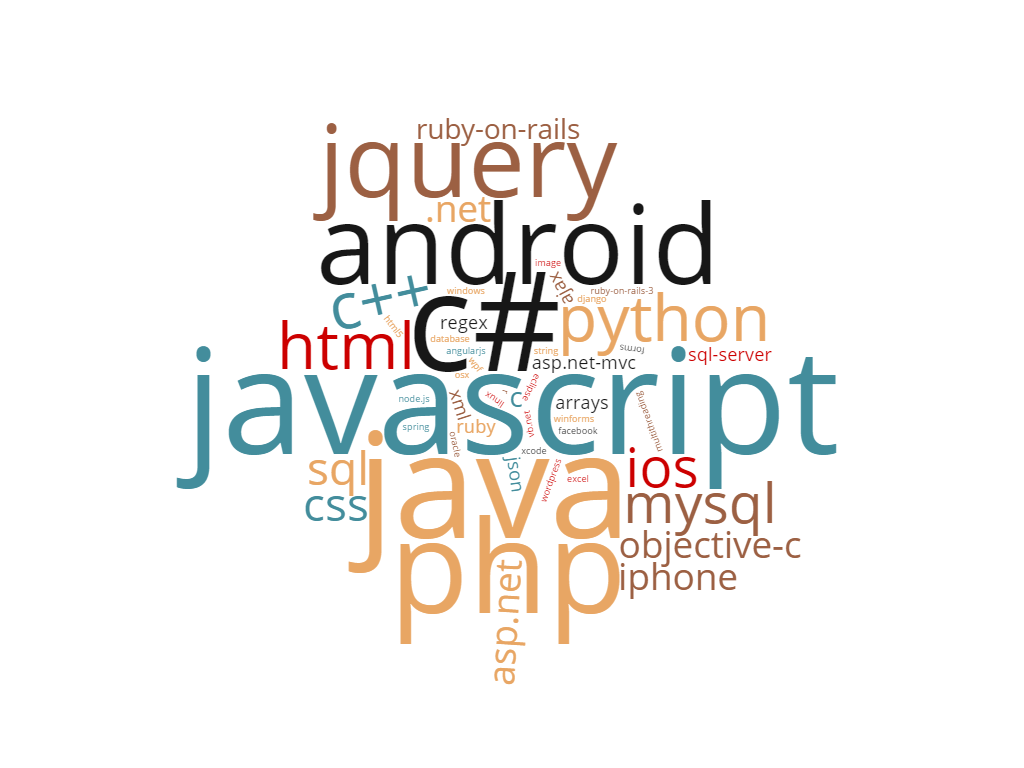}%
\label{fig:em_word_cloud}}
\hspace{0.1in}%
\subfloat[]{\includegraphics[width=2in]{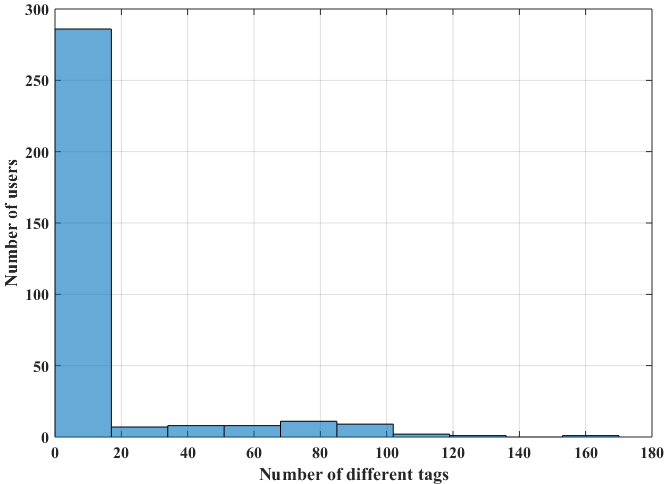}%
\label{fig:em_user_tag_freqs}}
\caption{ Visualization of tags of questions in our Stack Overflow dataset.
(a) Word cloud of tags, (b) Histogram of total number of tags used by users. A big portion of users only used fewer than 20 tags.
}
\label{fig:emerical_data_visualization}
\end{figure*}
\begin{table}[t]
\begin{center}
\caption{Selected Badges Information}
 \begin{tabular}{l l l}
 \hline
 Name & Description \\ 
 \hline\hline
 Curious & {Asked a well-received\footnote{A question is well-received if it is open and has a score greater than 0.
 \url{http://meta.stackexchange.com/questions/234259/asking-days-badges}} question on 5 separate days}& \\ 
 \hline
 Inquisitive & {Asked a well-received question on 30 separate days} & \\
 \hline
 Socratic & {Asked a well-received question on 100 separate days} &  \\
 \hline
 Explainer & Edited and answered 1 question (answer score $>$ 0) &  \\
 \hline
 Refiner & Edited and answered 50 question (answer score $>$ 0) &  \\
 \hline
 Refiner & Edited and answered 500 question (answer score $>$ 0) & \\
 \hline
\end{tabular}
\label{table:badge_stats} 
\end{center}
\end{table}

\begin{itemize}
\item \textbf {Synthetic Data.}
We considered $U=100$ users and $K=50$ different tags as the indicators of different topics. We also considered $|\mathcal{B}^q| = 5$ badges as threshold badges for asking questions, and $|\mathcal{B}^a| = 5$ badges as threshold badges for answering questions. The thresholds for badges are sampled from a uniform distribution $\tau_b \sim \mathbf{U}(0, 500)$. Then we generated random vectors of user interests ($\alpha_u$) and user expertise ($\eta_u$) over different tags by using a Dirichlet distribution, i.e. $\alpha_u, \eta_u \sim \mathbf{DIR}(50, 0.1)$.  
The exogenous intensities ($\mu_u^q, \mu_u^a$), and badge impact parameters ($\rho_u^q, \rho_u^a$) are drawn from uniform distributions, $\mu_u^q \sim \mathbf{U}(0, 0.01)$, $\mu_u^a \sim \mathbf{U}(0, 0.05)$, $(\rho_u^q, \rho_u^a) \sim \mathbf{U}(0,1)$.
Then by using the Ogata's thinning method \cite{ogata1981lewis}, we generated the events of our model. It is worth mentioning that we used both exponential ($g_w^{exp}(x,y)$) and Gaussian ($g_w^{gauss}(x,y)$) kernels for badge impacts. Using each kernel type,  we generated up to 1000 train events and 400 test events per user.  To eliminate the randomness in the results we repeated the procedure for 10 different times and aggregated the results.
\\

\item \textbf{Real Data.}
For the real dataset, we used the user questioning and answering data of Stack Overflow which is the most popular CQA website. To this end, we crawled the user data by using an API\footnote{\url{http://data.stackexchange.com/}}. We selected the 2000 top rank users of this website based on their reputation, and collected all their questions and answers. We selected 3 threshold badges for question activities, namely; \emph{Curious, Inquisitive, Socratic}. These badges are threshold badges that are awarded to the users based on the total active days for questioning. We also selected 3 threshold badges for answering activities; \emph{Explainer, Refiner,  Illuminator}. These badges are threshold badges that are awarded to the users based on total amount of their answers. 
The criteria for getting different badges are depicted in table \ref{table:badge_stats}.
These two types of badges have different criteria. The question badges are awarded based on active days, while the answer ones are awarded based on total amount of participation, which helped us to evaluate the generality of the proposed kernels for capturing the effect of badges.
We considered the first tag of each question as its mark. The word cloud of questions tags, and the histogram of total number of different tags used by different users is plotted in Fig. \ref{fig:emerical_data_visualization}. As it can be seen, a big portion of users, used fewer than $20$ different tags which is an indicator of limited interest of users in different domains.
It is worth mentioning that we only selected the events that participate in winning the selected badges. For the selected users and for each type of activities,  we considered 80\% of their actions as train data, and the remaining 20\% as the test data. 


\end{itemize}
\begin{figure*}[!t]
\centering
\subfloat[]{\includegraphics[width=2in]{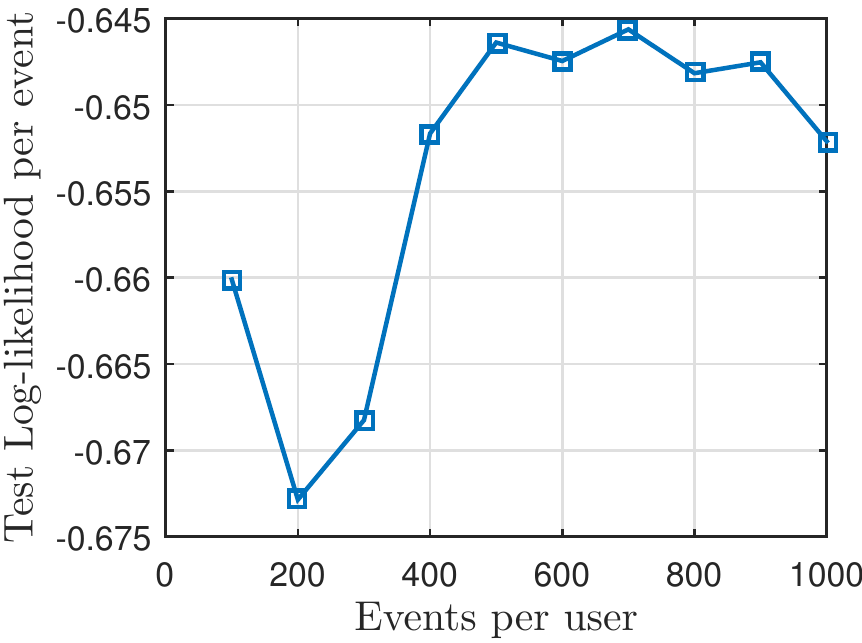}%
\label{fig:synt_overal_lglk}}
\hspace{0.1in}%
\subfloat[]{\includegraphics[width=2in]{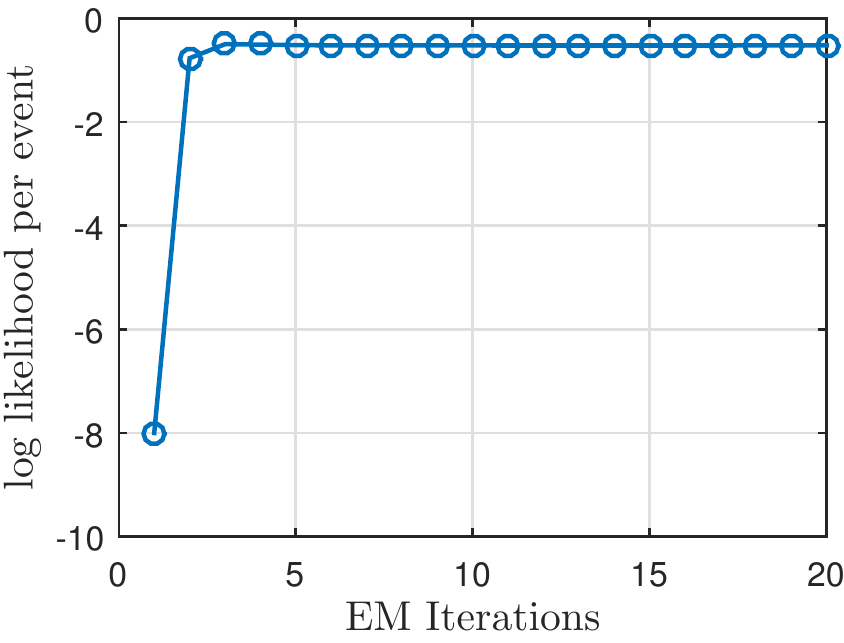}%
\label{fig:synt_em_lglk}}
\caption{ The performance of proposed inference algorithm in predicting the test data. The results are the average over question and answer data over 10 different runs.
(a) Per event log-likelihood of test data over different fractions of train data, (b) Per event log-likelihood over different iterations of EM algorithm.
}
\label{fig:synt_lglk}
\end{figure*}

\begin{figure*}[!t]
\centering
\subfloat[]{\includegraphics[width=2in]{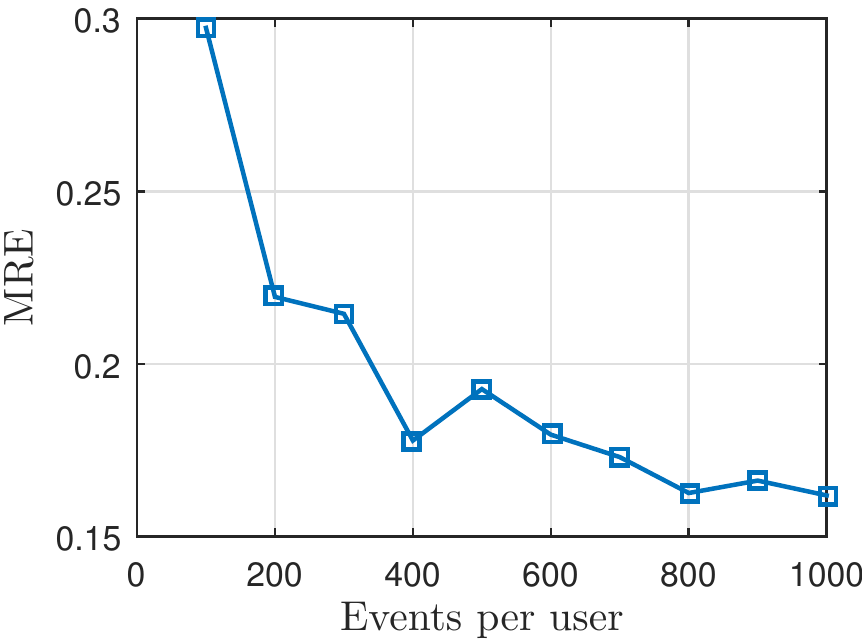}%
\label{fig:synt_mre_time_overal}}
\hspace{0.1in}%
\subfloat[]{\includegraphics[width=2in]{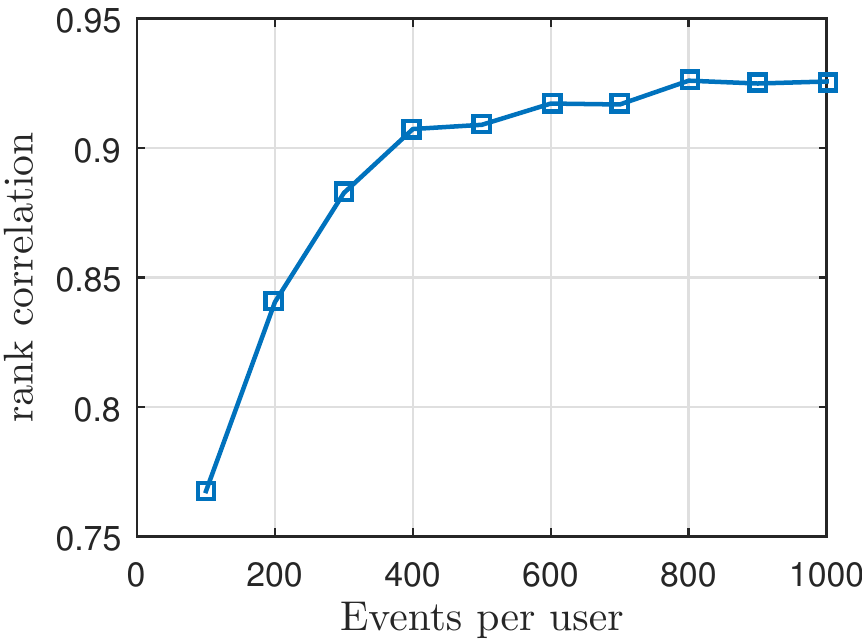}%
\label{fig:synt_rank_time_overal}}
\vfil
\subfloat[]{\includegraphics[width=2in]{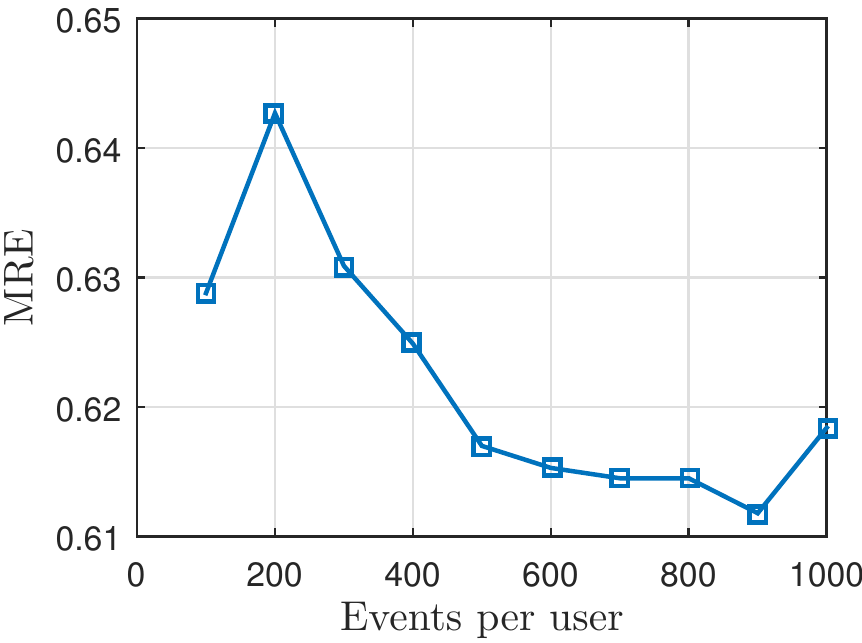}%
\label{fig:synt_mre_topic_overal}}
\hspace{0.1in}%
\subfloat[]{\includegraphics[width=2in]{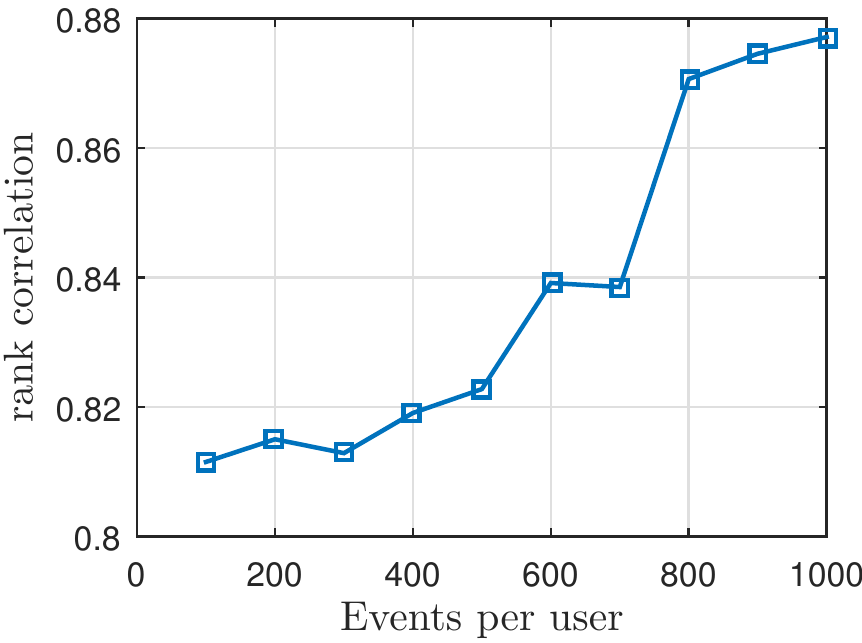}%
\label{fig:synt_rank_topic_overal}}
\caption{\texttt{MRE} and \texttt{Rank Correlation} of proposed method for recovering temporal and content parameters.
(a) Average \texttt{MRE} for temporal parameters vs the training size,
(b) Average \texttt{Rank Correlation} for temporal parameters vs the training size,
(c) Average \texttt{MRE} for content parameters vs the training size,
(d) Average \texttt{Rank Correlation} for content parameters vs the training size.
}
\label{fig:synt_errors}
\end{figure*}
\subsection{Synthetic Results}
We evaluated the performance of proposed inference algorithm by using the synthetic data. Indeed we are seeking to answer the following questions;
 1) \emph{What is the predictive performance of proposed algorithm?}, and, 2) \emph{Can the proposed inference algorithm accurately infer the model parameters?}

\subsubsection{What is the predictive performance of proposed algorithm?}
To evaluate the predictive performance of proposed method, we plotted the \texttt{Average log-likelihood per event} for the test data vs different train data sizes. Fig. \ref{fig:synt_overal_lglk} shows the results. We should note that the results are averaged over different kernel types and different users.
 As it can be seen, with the increase of training events, the performance of proposed method in predicting the future events increases. To evaluate the correctness of the variational EM algorithm, we also plotted average log-likelihood per event after the E-step vs the iteration number in Fig. \ref{fig:synt_em_lglk}. In the variational EM, log-likelihood after the E-step should never decreases \cite{Bishop2006PRM}, and our results conform with this expectation. It was also noticed that only after 4 steps the log-likelihood converged which is an indicator of fast convergence of the proposed inference algorithm.

\subsubsection{Can the proposed inference algorithm accurately infer the model parameters?}
To evaluate whether the proposed inference algorithm is able to accurately estimate the model parameters, we plotted the average \texttt{MRE} and \texttt{Rank} for the temporal and content parameters.  The \texttt{MRE} metric, measures the relative error between the true and estimated parameters. i.e. $\frac{1}{n} \sum_{i=1}^n \frac{|\theta_i - \hat{\theta}_i|}{\theta_i}$. We also evaluated how much the order of parameter values is preserved. To this end, we used the \emph{Kendall rank coefficient} \cite{kendallTau1938}. We averaged the metrics over all temporal parameters $\Theta_t = \left\lbrace\mu_u^{q}, \mu_u^a, \rho_u^q, \rho_u^a\right\rbrace_{u=1}^ {|U|}$ and content parameters $ \left\lbrace\alpha_u, \eta_u\right\rbrace_{u=1}^ {|U|}$. The average results over temporal and content parameters are depicted in Fig. \ref{fig:synt_errors}. 
We should note that the results are averaged over event sets of different kernel types and different users.
The first row (Fig. \ref{fig:synt_mre_time_overal}, \ref{fig:synt_rank_time_overal}) shows the average recovering error of temporal parameters for different train sizes, and the second row, (Fig. \ref{fig:synt_mre_topic_overal}, \ref{fig:synt_rank_topic_overal}) shows the average errors for content parameters. The results show that the performance of the proposed method in both metrics for both temporal and content parameters improves as the amount of train data increases.  The \texttt{MRE} decreases as the number of train data increases, and the \texttt{Rank Correlation} increases as the number of train data increases. These results confirms the superior performance of the proposed method.

\subsection{Real Data Results}
We also evaluated the performance of proposed method on real data gathered from Stack Overflow. We used the following criteria to evaluate the predictive performance of methods for time and mark predictions.
\begin{itemize}
\item \textbf{Average Test Log-Likelihood:} We calculated the average temporal log likelihood for test events of each user, and reported the average results over all users. The higher log likelihood means better performance.
\item \textbf{Time Prediction:}  For each user we predicted when the test events will occur using the density of next event times ($f(t) = \lambda(t)\exp{-\int_0^T\lambda(s) ds}$). To achieve this, we computed the expected time of next event by sampling the future events. We reported the Mean Absolute Error (\texttt{MAE}) and Mean Relative Error (\texttt{MRE}) between the predicted time and the true time.  
\item \textbf{Mark Prediction:} We also evaluated the predictive performance of the proposed method for predicting the marks of events.
As we mentioned before, the mark of a question is its tag, and the mark of an answer is the question it belongs to.
For each test event, given the time of event, we evaluated the probability of its mark (tag/parent). We rank all the tags/parents in the descending order of their probability, and create a recommendation list. We reported average \texttt{Precision@k} and \texttt{NDCG@k	}  \cite{charlin2015dynamic} for different test events over all users. \texttt{Precision@k} is the percent of predictions that the true value is among the top k predicted ones.
\end{itemize}
We also used two types of baselines for temporal and content evaluations. For temporal evaluations, measuring the effect of badges would help to obgtain predictions. We used the following baselines.
\begin{itemize}
\item \textbf{Hawkes Process:}  To see whether there is a self-excitation among the events. We fitted a simple Hawkes process for the events of each user with intensity  $\lambda_u^{q/a}(t) = \mu_u^{a/q} + \beta_u^{a/q} \sum_{e_i \in \mathbf{D}_u^{a/q}} \kappa(t, t_i)$, where $\kappa(t, t_i) = \exp(-(t-t_i))$.
\item \textbf{Homogeneous Poisson Process:}
We also fitted a Poisson Process with a constant intensity function for each user and type (question \& answer). $\lambda_u^{a/q} (t) =  \lambda_u^{a/q}$.
\end{itemize}
For mark evaluations, we used the following baselines.
\begin{itemize}
\item \textbf{Most Popular:} At each time step, regardless of the time we predicted the most popular marker. Usually, predicting the most popular is a strong heuristic \cite{du2015time}. Using this prediction, we can evaluate the impact of utilizing customization and time in our model. 
\item \textbf{Most Recent:} At each time step, we predict the most recent markers. Using this prediction, we can evaluate impact of utilizing the customization and preferences in our model.
\end{itemize}


\begin{figure*}[!t]
\centering
\subfloat[]{
\includegraphics[width=1.5in]{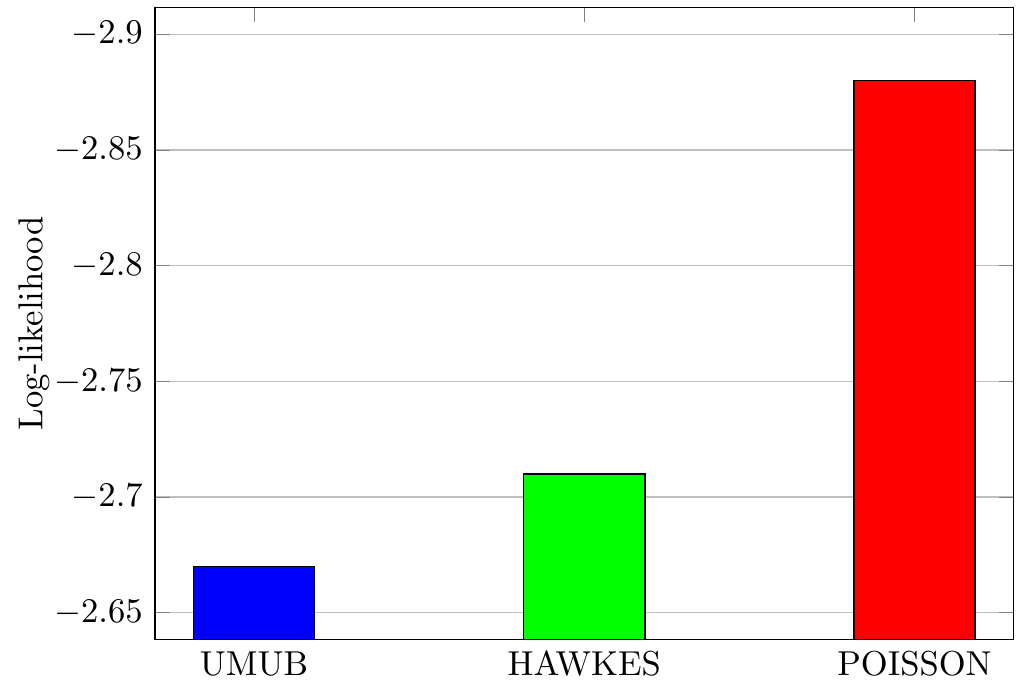}%
\label{fig:em_ans_lglk}}
\hspace{0.05in}%
\subfloat[]{\includegraphics[width=1.5in]{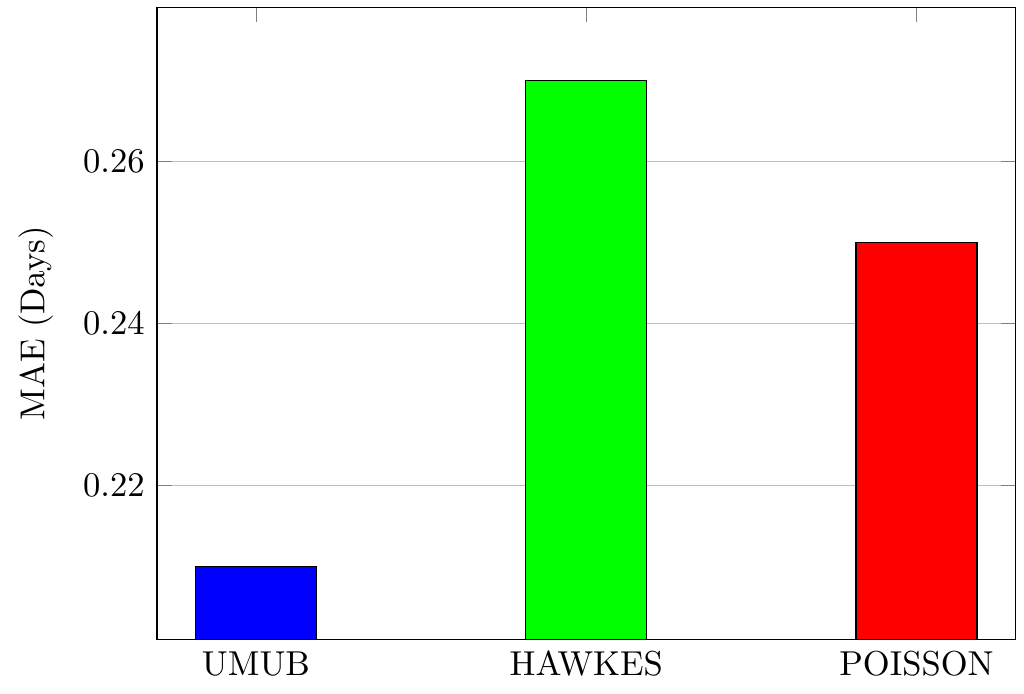}%
\label{fig:em_ans_mse}}
\hspace{0.05in}%
\subfloat[]{\includegraphics[width=1.5in]{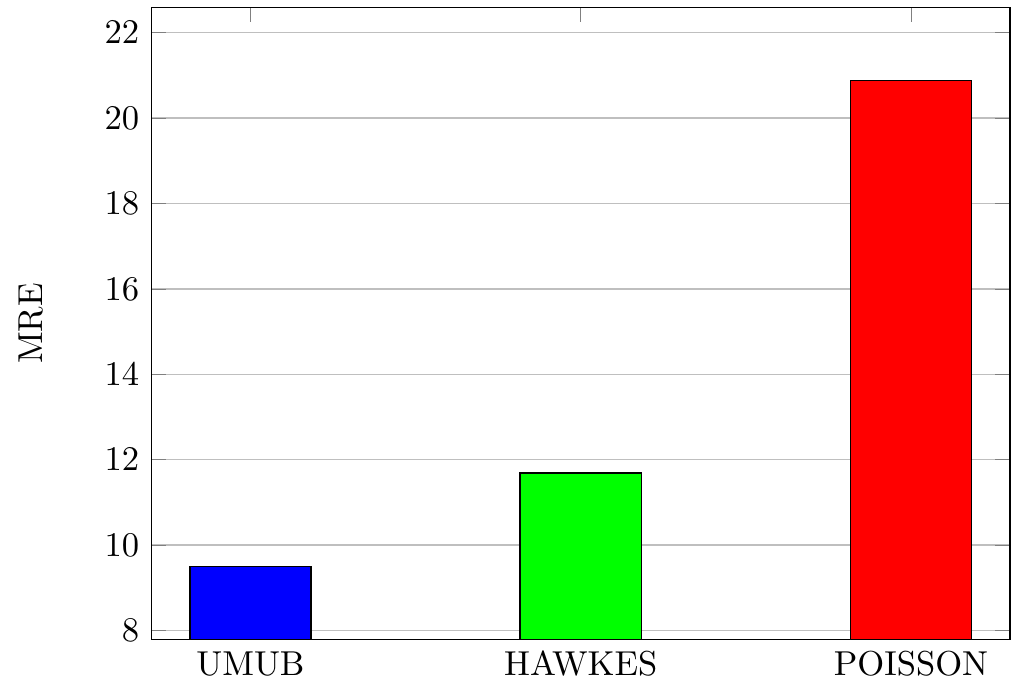}%
\label{fig:em_ans_mre}}
\vspace{0.05in}
\subfloat[]{\includegraphics[width=1.5in]{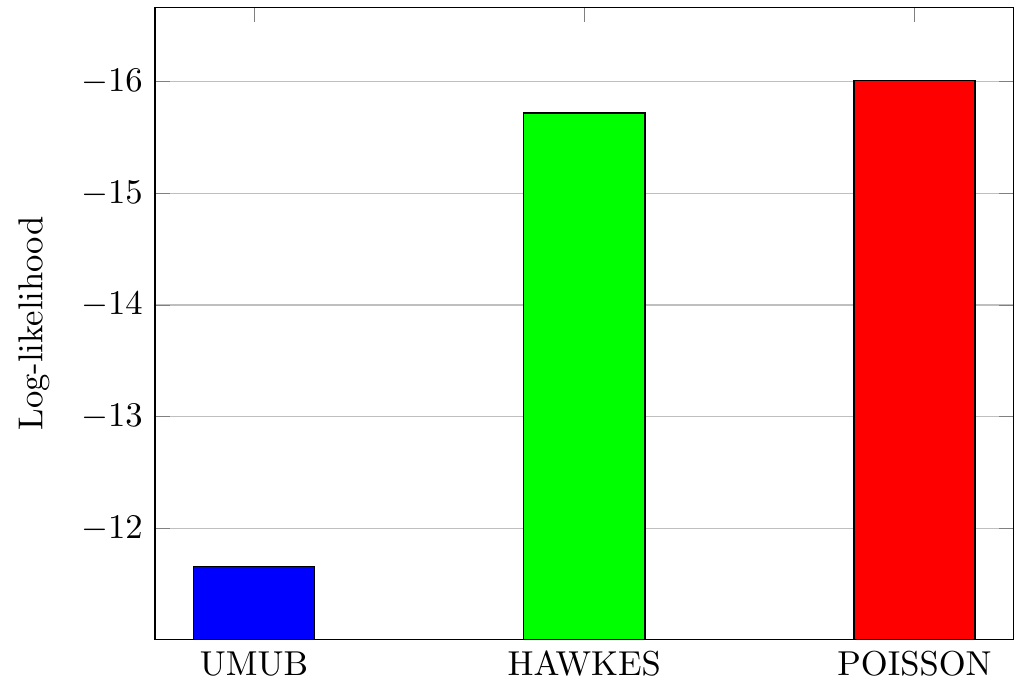}%
\label{fig:em_quest_lglk}}
\hspace{0.05in}%
\subfloat[]{\includegraphics[width=1.5in]{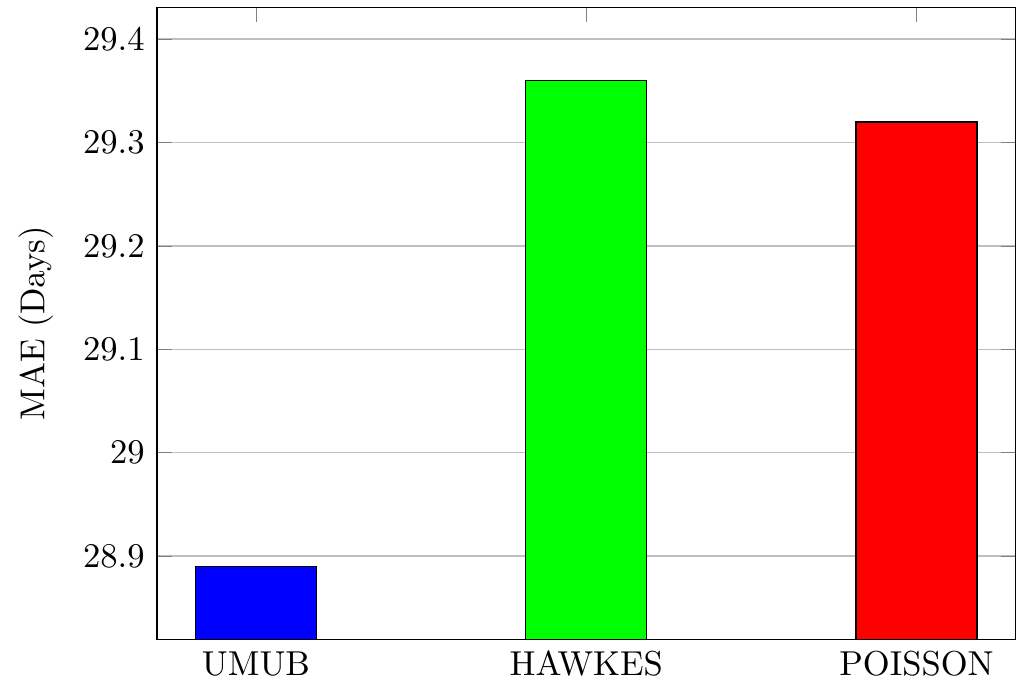}%
\label{fig:em_quest_mse}}
\hspace{0.05in}%
\subfloat[]{\includegraphics[width=1.5in]{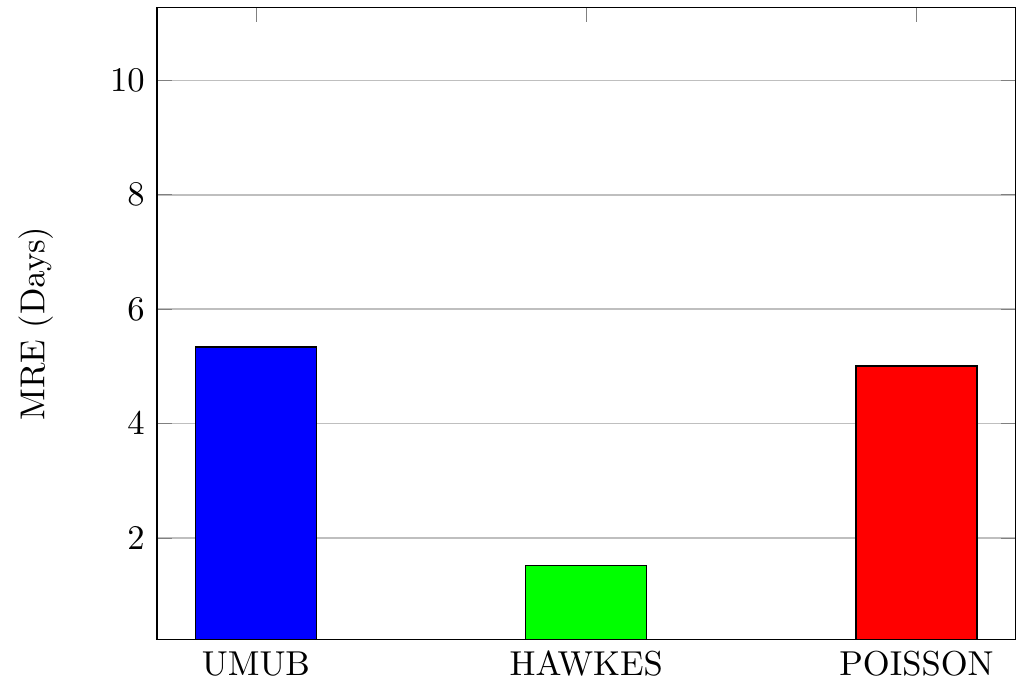}%
\label{fig:em_quest_mre}}
\hspace{0.05in}%
\subfloat[]{
\includegraphics[width=2in]{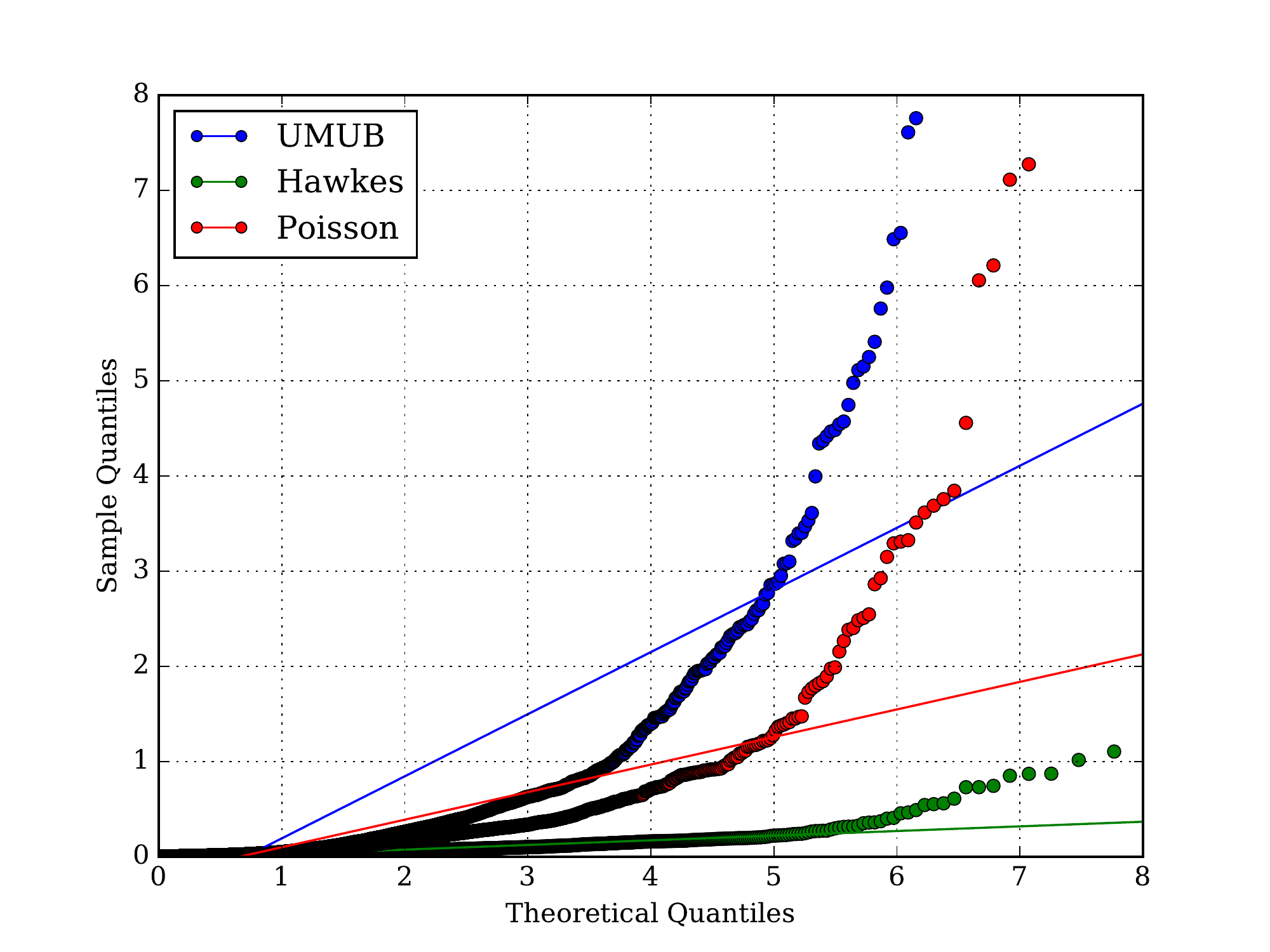}%
\label{fig:em_ans_qqplot}}
\hspace{0.05in}%
\subfloat[]{
\includegraphics[width=2in]{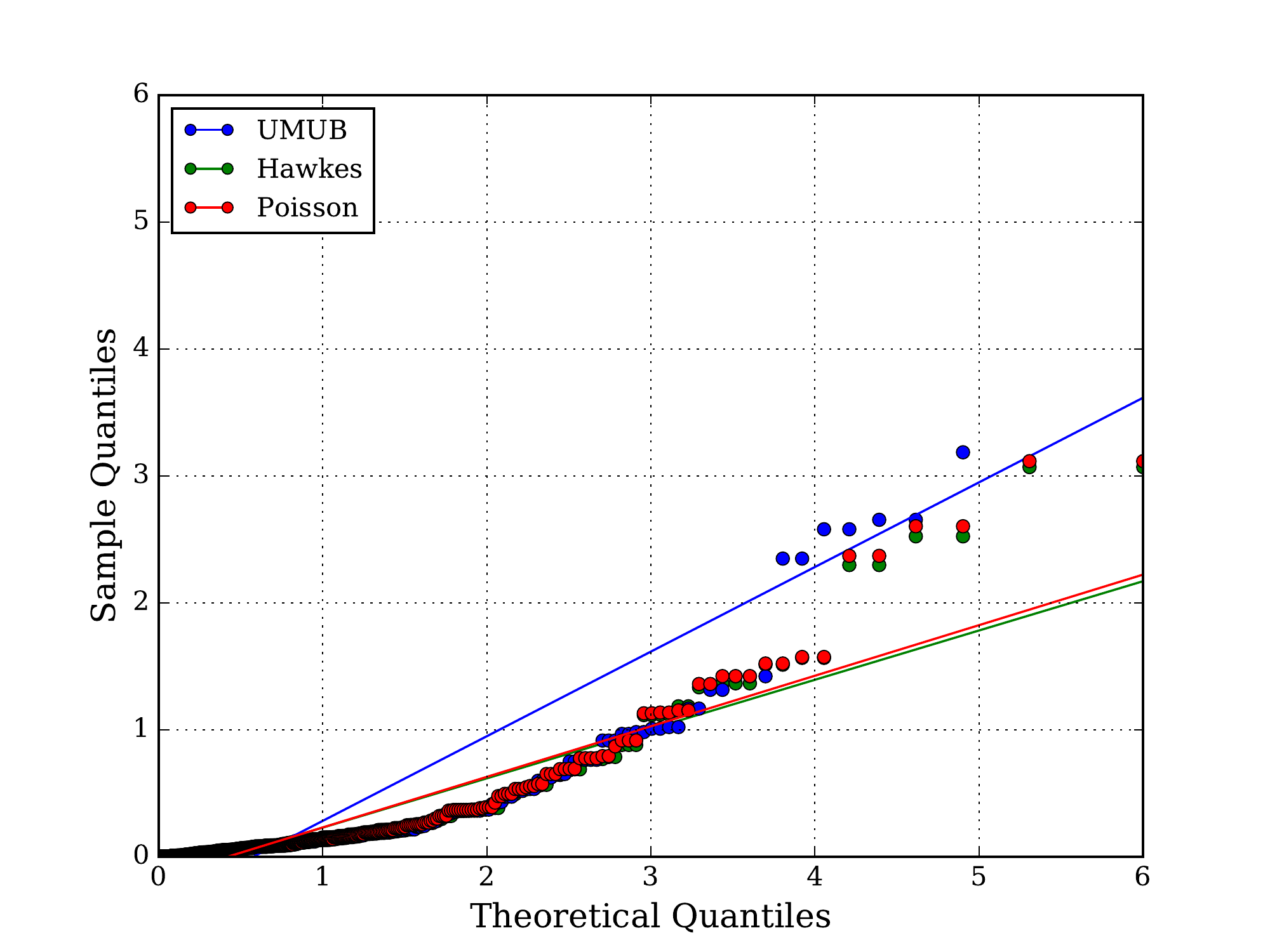}%
\label{fig:em_quest_qqplot}}
\caption{ Real data temporal results. 
First row shows the average (a) log likelihood, (b) MAE, (c) MRE for test answer events over all users.
Second row depicts the average (a) log likelihood, (b)MAE, (c)MRE for test question events over all users.
}
\label{fig:temporal_error_results}
\end{figure*}
\subsubsection{Temporal Results:}
In Fig. \ref{fig:temporal_error_results}, we have plotted the performance of different methods in predicting the time of next events for both question and answer events. For the log-likelihood metric, (Figs. \ref{fig:em_quest_lglk},\ref{fig:em_ans_lglk}), the proposed method outperforms the Hawkes and Poisson precesses. As it can be seen, the performance of proposed method degrades from answers to questions, because asking more questions is a hard task, and the impact of badges on asking more questions is less than their impact on answering. Therefore, the performance of proposed method degrades from answering to asking questions.  For the other two metrics, the proposed method also outperforms the competitors. Only for the \texttt{MRE} metric, the results of Hawkes are better. This is because the Hawkes method predicted times are usually smaller than the true times, which result in smaller \texttt{MREs}. It is interesting to note that the results of Hawkes and Poisson processes are similar for asking questions. This is because asking questions is usually driven by external factors rather than the history of events. Therefore, their performances will be closer for asking questions. 
According to the time-change theorem \cite{daley2008II}, given all $t_i$ and $t_{i+1}$ subsequent event times of a particular point process, the set of intensity integrals $\int_{t_i}^{t_{i+1}}\lambda(t)dt$ should conform to the unit-rate exponential distribution, if the samples are truly sampled from the process $\lambda(t)$. Hence, we compare the theoretical quantiles from the exponential distribution with the ones from different models to the real sequence of events. The closer the slope is to one, the better a model matches the event patterns. Figs. \ref{fig:em_ans_qqplot}, \ref{fig:em_quest_qqplot} show the results for different models for answers and questions, respectively. The results show that our UMUB model can better explain the observed data compared to the other models.

\begin{figure*}[!t]	
\centering
\subfloat[]{\includegraphics[width=2in]{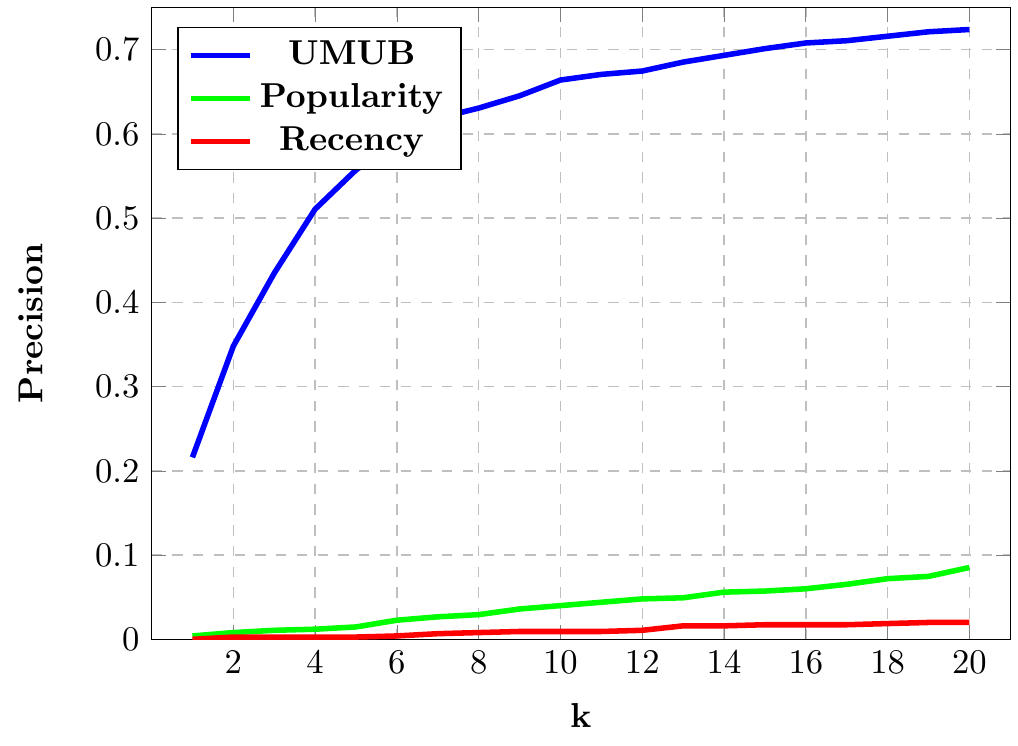}%
\label{fig:answer_precision_at_k}}
\hspace{0.1in}%
\subfloat[]{\includegraphics[width=2in]{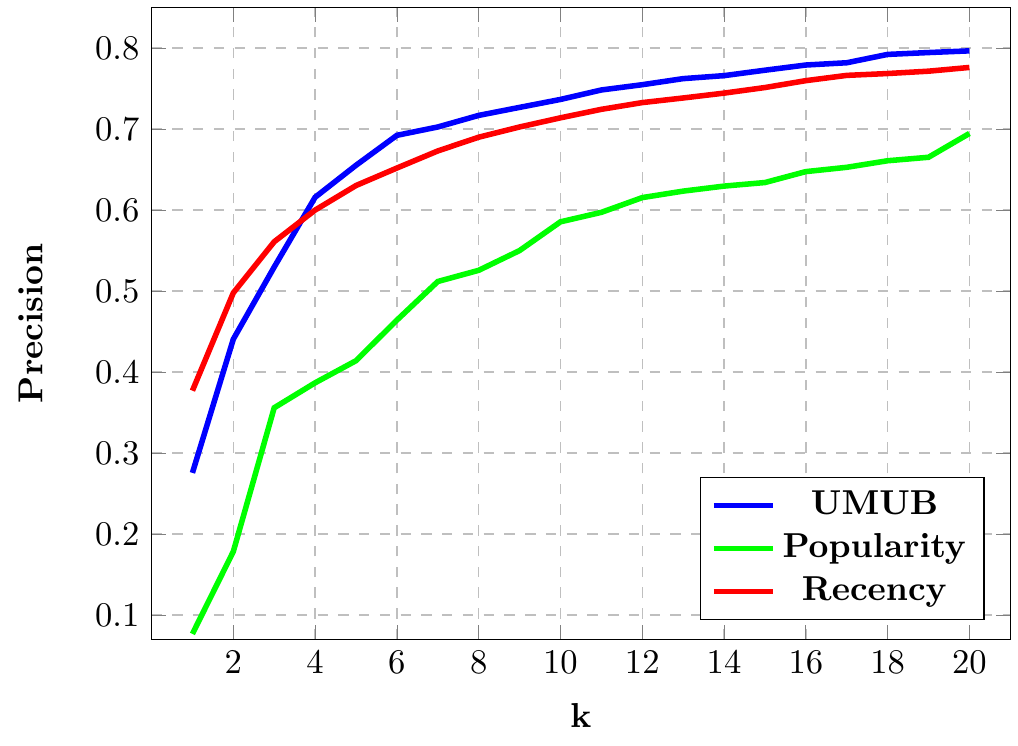}%
\label{fig:question_precision_at_k}}
\hspace{0.1in}%
\subfloat[]{\includegraphics[width=2in]{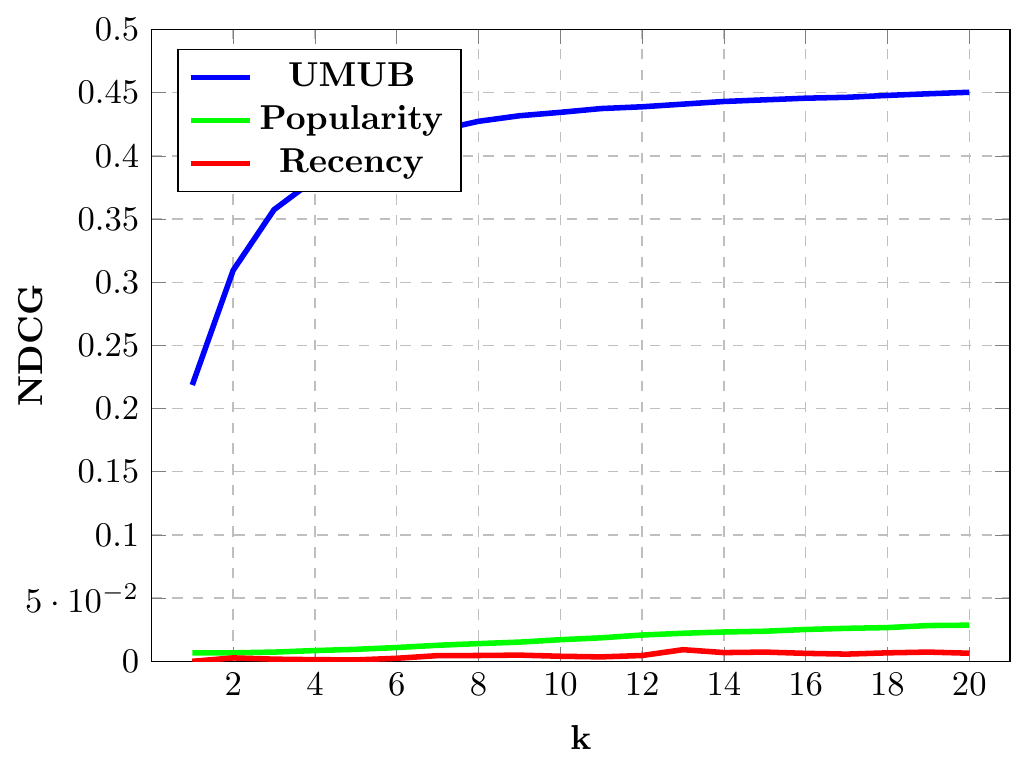}%
\label{fig:answer_ndcg_at_k}}
\hspace{0.1in}%
\subfloat[]{\includegraphics[width=2in]{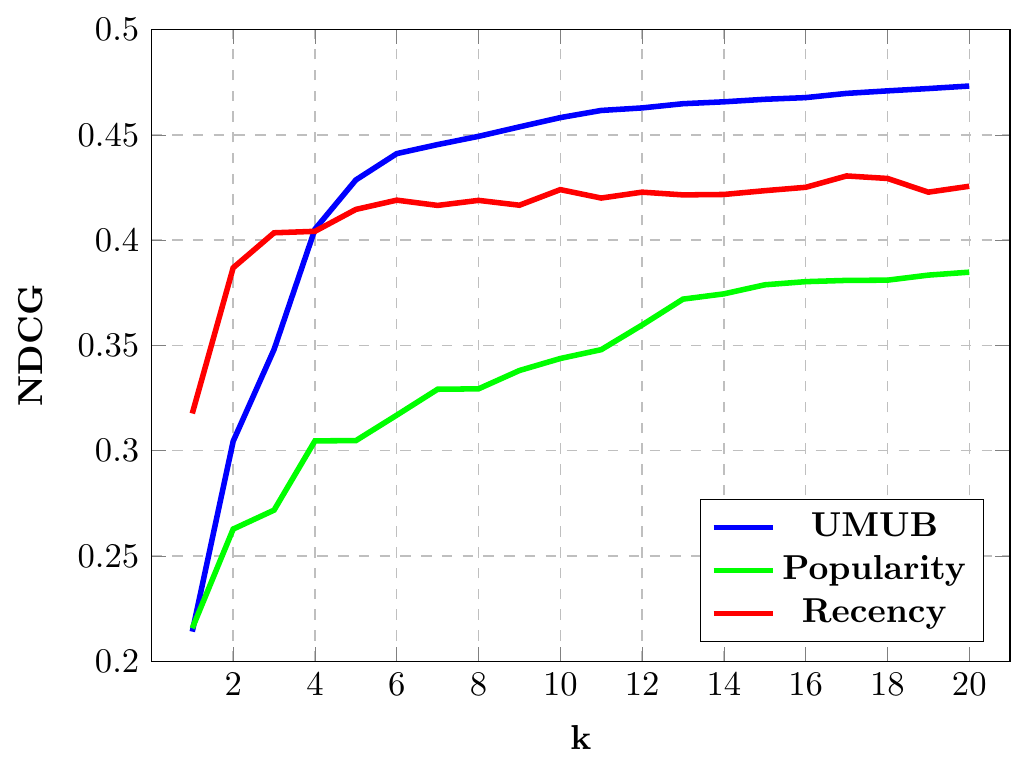}%
\label{fig:question_ndcg_at_k}}

\caption{ The results for predicting the mark of events.
Average \texttt{ precision@k} (a) and \texttt{NDCG@k} (c) for predicting the parent of answers over all users, for $k=1$ to $20$. 
Average \texttt{ precision@k} (b) and \texttt{NDCG@k} (d) for predicting the tag of questions over all users, for $k=1$ to $20$.
}
\label{fig:mark_em_results}
\end{figure*}

\subsubsection{Mark Prediction Results}
We plotted the results of different methods in predicting the mark of test events in Fig. \ref{fig:mark_em_results}. Fig. \ref{fig:answer_precision_at_k} and Fig. \ref{fig:answer_ndcg_at_k}, show the \texttt{ precision@k} and \texttt{NDCG@k} for different methods in predicting the parent of answers versus different $k$. In the same way, Fig. \ref{fig:question_precision_at_k} and \ref{fig:question_ndcg_at_k} show the \texttt{ precision@k} and \texttt{NDCG@k} for different methods in predicting the tag of questions versus $k$.
As it can be seen, by increasing $k$, the performance of all methods improves. The proposed method outperforms the competitors in predicting both the parents and tags.
For predicting the tag of questions, the most recent method performs close to the proposed method (Figs. \ref{fig:question_precision_at_k}, \ref{fig:question_ndcg_at_k}). This is because the asking questions is a sequential task, and usually a user asks questions in a domain, sequencially. In addition, by the fact that users usually ask questions in a limited number of topics ( Fig. \ref{fig:em_user_tag_freqs}), the recency is a strong predictor for the tags of questions.
The high difference between the proposed method and the competitors for predicting the parent of answers (Figs. \ref{fig:answer_precision_at_k},  \ref{fig:answer_ndcg_at_k}) is because, although the popularity and recency heuristics are strong predictors, but for predicting the questions that the user will answer they can not work well. This is due to the huge amount of questions that arrive in every moment and the user must select one of them. Therefore, paying attention only to time or popularity will result in poor performance. The proposed method, considers both the personal preferences and the temporal impacts in predictions, and hence it can better predict the question that user will answer.

\begin{figure*}[!t]
\centering
\subfloat[]{\includegraphics[width=2in]{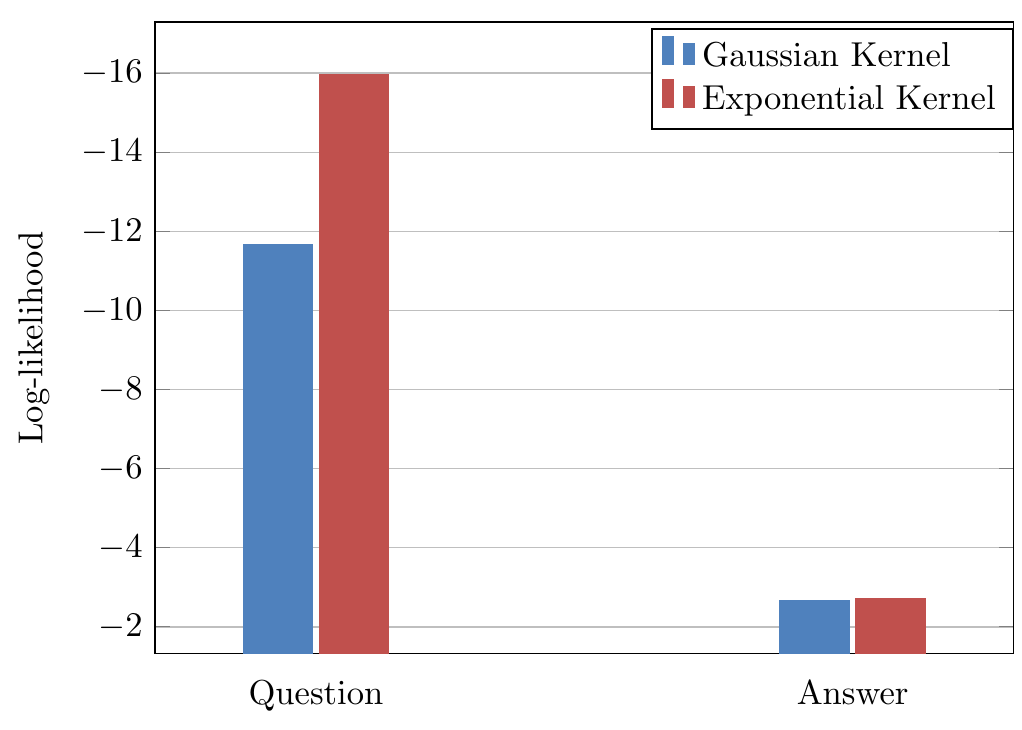}%
\label{fig:em_impact_of_kernels}}
\hspace{0.1in}%
\subfloat[]{\includegraphics[width=2in]{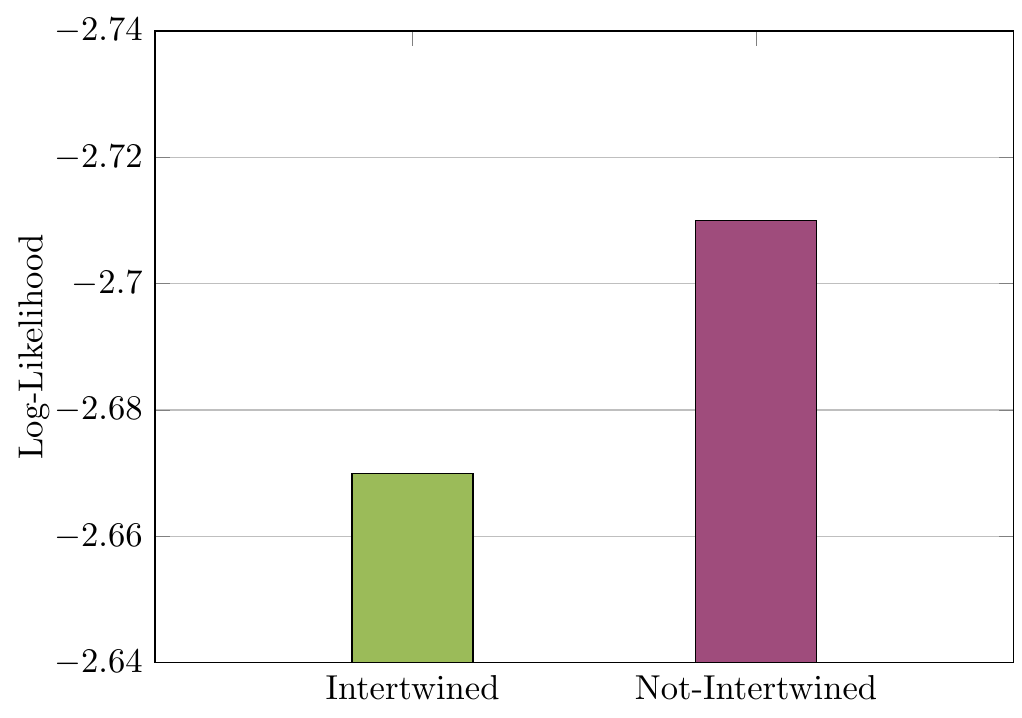}%
\label{fig:answer_impact_of_hist}}
\caption{ The impact of different configurations on the performance of proposed method.
(a) Comparison of the performance of Gaussian and Exponential Kernels on both asking questions and answering processes.
(b) The impact of questions history on the performance of proposed method for answering questions.
}
\label{fig:em_kernels_history_impact}
\end{figure*}
\subsubsection{Impact of Kernels and Intertwining the Processes}
We also studied the impact of different configurations on the performance of the proposed method. 
First, we studied the impact of different kernels on the performance.
Fig. \ref{fig:em_impact_of_kernels} shows the impact of \textit{Gaussian} and \textit{Exponential} kernels, which  previously we depicted in Fig. \ref{fig:kernels}, on the Log-likelihood of test events. The results shows that the Gaussian kernel works better than exponential for both type of actions. This means that for both questioning and answering actions, the impact of badges on user actions do not drop suddenly after obtaining the badge. In addition, the badges have an impact on user activities even after obtaining them. 

We also studied the impact of intertwining the processes on the performance of the proposed method. We compared the two versions of proposed method with considering the impact of questions on answers and without considering the effect. Fig. \ref{fig:answer_impact_of_hist} shows the performance of the two versions based on the Log-Likelihood of  test events. As it can be seen, the intertwined version performs better than the version without the impact of questions. This, shows that the questioning process have an impact, though little, on the answering process.

%

%

\section{Conclusion}\label{sec:conclusion}
In this  paper, we proposed new continuous-time user models by using a powerful mathematical framework; Temporal Point Processes.  Unlike, previous works on continuous-time user modeling which mainly focus on the impact of peer influence on user actions, the proposed method also consideres the impact of content and gamification elements, especially badges, on user actions. We extended the proposed method for modeling user questioning and answering activities over CQA cites and proposed an inference algorithm based on Variational EM which can efficiently learn the model parameters. The learnt parameters can help in categorizing users and understanding their preferences. These information will help the social media owners in creating implicit user profiles and  delivering customized services. The empirical evaluations on both synthetic and real datasets, demonstrate the superior predictive performance of the proposed method.	

There are many interesting lines for future works. In this work, we only considered the single activity threshold badges, incorporating other hybrid badges and also other gamification elements like reputation systems is a new line of research. On the content side, we used simple models, considering more complex models such as Bayesian non-parametric ones will help to improve the model. Modeling the quality of user actions and considering the impact of badges on the quality, is another interesting line of future investigation.

%
%


\bibliographystyle{spmpsci}      
\bibliography{ref}   


\end{document}